\documentstyle[12pt,rotating]{article}
\newcommand{\Frac}[2]{\frac{\displaystyle #1}{\displaystyle #2}}

\newcommand{\ra}{\rangle}
\newcommand{\la}{\langle}
\newcommand{\opc}{${\cal O}(p^4)$ }
\newcommand{\ops}{${\cal O}(p^6)$ }
\newcommand{\pra}{${\cal A}$ }
\newcommand{\prb}{${\cal B}$ }

\def\mapright#1{\smash{
     \mathop{\longrightarrow}\limits^{#1}}}
\begin{document}
\pagestyle{empty}
\begin{titlepage}
\begin{center}
\vspace*{-1.1cm}
\hfill  INFNNA-IV-97/27\\
\hfill  DSFNA-IV-97/27\\
\hfill {\tt hep-ph/9711211} \\
\hfill October, 1997
\vspace*{1.4cm} \\
{\LARGE \bf  Spin-1 resonance  contributions\\ 
   to the weak Chiral Lagrangian:   \\the vector field formulation ${}^*$\\ }
\vspace*{1.6cm}
{\large \sc Giancarlo D'Ambrosio$^{\dagger}$} $ \; \; $ and $ \; \; $
{\large \sc  Jorge Portol\'es$^{\ddagger}$ }
\vspace*{0.4cm} \\
Istituto Nazionale di Fisica Nucleare, Sezione di Napoli \\
Dipartamento di Scienze Fisiche, Universit\`a di Napoli \\
I-80125 Napoli, Italy \\ 
\vspace*{1.8cm} 

\begin{abstract}
We use the Vector formulation to evaluate vector and axial--vector exchange
contributions to the  ${\cal O}(p^4)$ weak Chiral Lagrangian.
We  recover in this framework the bulk of the contributions found
previously by Ecker et al. in the antisymmetric formulation of vectors and
axial--vectors, but new interesting features arise~:
i) most of our results are independent of Factorization and
ii) novel contributions to non-leptonic kaon decays, proper of this 
formulation and phenomenologically interesting,  are found.
The phenomenological implications for $K \rightarrow \pi\pi(\pi)$ and 
radiative (anomalous and non-anomalous) non--leptonic kaon decays 
are thus investigated  and found particularly relevant. 
\end{abstract}
\end{center}
\vspace*{0.4cm}
PACS~: 12.15.-y,12.39.Fe,12.40.Vv,13.25.Es \\
Keywords~: Radiative non--leptonic kaon decays, Non--leptonic weak
hamiltonian, Chiral Perturbation Theory, Vector meson dominance.
\vspace*{0.4cm}\\
$^{\dagger}$ E-mail~: dambrosio@axpna1.na.infn.it \\
$^{\ddagger}$ E-mail~: portoles@axpna1.na.infn.it \\
\vfill
\noindent * Work supported in part by HCM, EEC--Contract No. 
CHRX--CT920026 (EURODA$\Phi$NE).
\end{titlepage}
\newpage
\pagestyle{plain}
\pagenumbering{arabic}

\section{Introduction}
\hspace*{0.5cm}
Kaon decays are an important laboratory 
\cite{DE95,reviews,EDU95,CGE95,TONI95,DAGI96}
to understand weak interactions, chiral dynamics and CP violation.
Present and future experiments  will definitely improve our 
knowledge in this area \cite{Orsay}. The natural framework to study 
kaon decays is Chiral Perturbation Theory ($\chi$PT) \cite{WE79,GL85,MG84}, 
where matrix elements, consistent with  chiral symmetry and its 
spontaneous breaking ($SU(3)_L \otimes SU(3)_R \rightarrow
SU(3)_V$ plus eight Goldstone bosons to be identified with the octet of
lightest pseudoscalar mesons), are written as a perturbative expansion 
in masses and external momenta. At leading ${\cal O}(p^2)$ 
current algebra features like PCAC are recovered. 
\par 
At next \opc \cite{WE79,GL85} a loop expansion, required by unitarity,
and local operators, consistent with  chiral symmetry, appear.
The divergent part of the coefficients
of these local operators are needed to reabsorb divergences
in the loop contributions and thus are called counterterms.
The finite part of a counterterm has to be determined
from the phenomenology or through extra theoretical assumptions \cite{EDU95}
(and references therein).
\par 
Vector Meson Dominance (VMD) has proven to be very efficient
in describing the coefficients of the \opc strong lagrangian 
\cite{EG89,DR89,BK88}, giving a good agreement with the phenomenology. 
However VMD in the strong sector is implemented automatically at \opc 
only by the antisymmetric formulation of the vectors \cite{EGL89}
while in the usual vector formulation has to be just imposed. This is 
not necessarily the case at higher chiral orders where the
conventional vector formulation is able to recover
structures required by QCD that the antisymmetric formulation
does not generate \cite{EP90}.  
\par 
We will be interested here in the study of the spin--1 resonance 
contributions to non--leptonic kaon decays in the $\chi$PT framework.
Some of these processes receive contributions already at ${\cal O}(p^2)$
(that are not generated by resonance exchange) with sizable \opc corrections, 
like $K\rightarrow\pi\pi$ and $K\rightarrow\pi\pi\pi$.
Others start at \opc, like $K\rightarrow\pi\gamma^\ast$,
$K\rightarrow\pi\gamma\gamma$, direct emission in 
$K\rightarrow\pi\pi\gamma$, ... The full weak \opc
counterterm structure has been analysed \cite{KM90,EK93,Esposito90}
and there are already interesting counterterm coefficient relations to be 
considered \cite{DE95,DAGI96,DE97,isidori96}.
The question of the dominance of vector meson exchange 
in weak decays has not yet a clear answer due to the unknown weak couplings 
of vector and axial--vector mesons. Thus one has to rely on models to make 
quantitative predictions.
The expansion in inverse power of number of colours
($1/N_c$) has motivated several researches \cite{EDU95,PI91} (and references
therein) and their consequences for the \opc weak chiral lagrangian
have been studied by direct evaluation of the full
short distance weak hamiltonian in the large $N_c$ 
limit and then bosonizing in terms of the lightest degrees of freedom i.e. 
the Golstone boson fields \cite{BP93}. 
The general structure of the \opc weak chiral lagrangian generated
by resonance  exchange has been studied in Ref. \cite{EK93}
where vectors and axial--vectors are implemented in the antisymmetric
formulation and then the Factorization Model (FM), justified by $1/N_c$ 
arguments, has been used. Several predictions and useful relations have 
been found in this way \cite{EK93}.
\par
The questions we address here are:
\begin{itemize}
\item[i)] Do these relations depend on the vector realization?
What about if we use the conventional vector formulation instead of the
antisymmetric one?
\item[ii)] What is the relative r\^ole of the  Factorization 
hypothesis in the two formulations?
\end{itemize}
We will see that the answers provide new results relevant for the 
phenomenology of most of the processes involved. Moreover we will 
conclude that the model dependence (factorization) of the results is 
very much suppressed if the vector formulation of the resonance fields
is used.
\par
The scheme of the paper is the following. In Section 2 we review
the $\chi$PT formalism in its strong and weak sectors. We also collect
and analyse critically the main results of previous work on this topic
that we would like to compare with. In Section 3 we deduce the \opc
weak chiral lagrangian as generated by vector and axial--vector 
contributions in the conventional vector formulation. We compare our 
results with the ones achieved by the antisymmetric formulation in 
Ref. \cite{EK93}. In Section 4 the relevant phenomenological implications
of our new results are pointed out. We report our conclusions in Section 
5. A brief appendix complements the main text.

\section{Chiral Perturbation Theory and weak interactions}
\hspace*{0.5cm}
In this Section we first review the formulation of $\chi$PT in the 
treatment of the strong and weak interactions. We also analyse the
previous results of the spin--1 resonance exchange generated 
\opc weak Chiral Lagrangian using the Factorization Model when the
antisymmetric field formulation of the resonance fields is implemented.

\subsection{Chiral Perturbation Theory} 
\hspace*{0.5cm}
$\chi$PT \cite{WE79,GL85} is the effective quantum field theory for the study
of low energy strong interacting processes. It relies in the exact 
$G \equiv SU(3)_L \otimes SU(3)_R$ global chiral symmetry of massless QCD. 
This symmetry group is assumed to break spontaneously to 
$H \equiv SU(3)_{L+R = V}$ generating an octet of Goldstone bosons
that are identified with the lightest octet of pseudoscalar mesons. 
Following Ref. \cite{CCW69} it turns out convenient to introduce
the Goldstone bosons $\varphi_i$, $i = 1, ...8$ through the $SU(3)$ matrix
$u(\varphi)$, which also parameterizes the coset space $G/H$ as 
\begin{equation}
u (\varphi) \, = \, \exp \left( \, \Frac{i}{2F} \sum_{j=1}^8 \, \lambda_j 
\varphi_j \, \right) \; \; \; \; , 
\label{eq:uphi}
\end{equation}
where $\lambda_i$ are the $SU(3)$ Gell--Mann matrices \footnote{Normalized
to $Tr(\lambda_i \lambda_j) = 2 \delta_{ij}$.} and $F \, \sim \, F_{\pi} 
\, \simeq \, 93 \, \mbox{MeV}$ is the decay constant of the pion. The
transformation under $(g_L,g_R)$ $\in G$ is 
\begin{equation}
u(\varphi) \, \mapright{G} \, g_R \, u(\varphi) \, h(g,\varphi)^{\dagger}
\, = \, h(g,\varphi) \, u(\varphi) \, g_L^{\dagger} \; \; \; , 
\label{eq:urht}
\end{equation}
where the compensator field $h(g,\varphi) \, \in \, H$ has been introduced.
Green functions and symmetry breaking terms can be more easily generated
by promoting the global symmetry to a local one, introducing external
fields. A covariant derivative on the $U(\varphi) \, = \, uu$ field is then 
defined as
\begin{equation}
D_{\mu} U \, = \, \partial_{\mu} U \, 
- \, i r_{\mu} U \, + \, i U \ell_{\mu} \; \; \; ,
\label{eq:ducov}
\end{equation}
where $\ell_{\mu} \, = \, v_{\mu} \, - \, a_{\mu}$ and $r_{\mu} \, = \, 
v_{\mu} \, + \, a_{\mu}$, are the left and right external fields, respectively,
in terms of the external vector and axial fields. 
\par
The leading ${\cal O} (p^2)$ strong lagrangian is
\begin{equation}
{\cal L}_2 \, = \, \Frac{F^2}{4} \, \langle \, u_{\mu} \, 
u^{\mu} \, + \, \chi_{+} \,  \rangle \; \; \; , 
\label{eq:str2}
\end{equation}
where
\begin{eqnarray}
u_{\mu} \, & = & \, i \, u^{\dagger} \, D_{\mu} \, U \, u^{\dagger}~,
\nonumber \\
\chi_{\pm} \, & = & \, u^{\dagger} \, \chi \, u^{\dagger} \, \pm \, 
u \, \chi^{\dagger} \, u~, \nonumber \\  
\chi \, & = & \, 2 \, B_{\circ} \, ( s \, + \, i \, p) \, = \, 
2 \, B_{\circ} \, {\cal M} \, + ... \, ,    
\label{eq:extrasym}  \\
{\cal M} \, & = & \, diag(m_u \, , \, m_d \, , \, m_s)~, \nonumber \\
 B_{\circ} \, & = &  \, - \, \Frac{1}{F^2} \, \langle 0 | \,
\overline{u} u \, |0 \rangle~, \nonumber 
\end{eqnarray}
and $\langle \, A \, \rangle  \, \equiv \, Tr(A)$ in the flavour space.
In the definition of $\chi$, $s$ and $p$ are the scalar and pseudoscalar 
external fields.
\par
Starting at \opc the strong chiral lagrangian has two well differentiated
components corresponding to vertices generating even-- and 
odd--intrinsic parity transitions. The \opc even--intrinsic parity lagrangian
${\cal L}_4$ was developed in Ref. \cite{GL85} and can be written as
\begin{equation}
{\cal L}_4 \, = \,  
\sum_{i=1}^{10} \, L_i \, O_i \; + \, H_1 \, O_{11} \; +\,H_2 \, O_{12}\;
 \; \; \; \; \; . 
\label{eq:strong4}
\end{equation}
Here $O_i$, $i=1,...12$ are local operators in terms of the pseudoscalar
and external fields. The couplings $L_i$ are rather well determined 
phenomenologically while $H_1$ and $H_2$ are not because the associate
operators only involve external fields and therefore do not contribute
to low--energy processes with pseudoscalars.
\par
The \opc odd--intrinsic parity lagrangian arises as a solution to the
Ward condition imposed by the chiral anomaly \cite{WZW71}. 
\par
The inclusion of other quantum fields than the pseudoscalar Goldstone bosons
in the chiral lagrangian was also considered in Ref.~\cite{CCW69}. We are 
interested in the introduction of vector and axial--vector resonances 
coupled to the non-linear realization of the Goldstone bosons
$u(\varphi)$ and to the external fields. 
It is well known that the incorporation of spin--1 mesons in the chiral
lagrangian is not unique and several realizations of the field can be
employed \cite{BI96}. In particular the antisymmetric formulation of vector
fields was seen to implement automatically vector meson dominance at \opc
in $\chi$PT \cite{EG89}. In Ref.~\cite{EGL89} was shown that, at \opc in 
$\chi$PT, once high energy QCD constraints are taken into account, the 
usual realizations (antisymmetric, vector, Yang--Mills and Hidden 
formulations) give equivalent low--energy effective actions. Although 
the antisymmetric tensor formulation of spin--1 mesons was proven to have
a better high--energy behaviour than the vector field realization at \opc
this fact is not necessarily the case at higher orders. In fact for the
odd--intrinsic parity action relevant in $V \rightarrow P \gamma$ decays
the antisymmetric tensor formulation gives a leading contribution at \opc
while QCD requires explicit ${\cal O}(p^3)$ terms that are provided by 
the vector formulation \cite{EP90}. The analogous situation in the weak
sector has not been studied yet. The authors of Ref.~\cite{EK93} used the
antisymmetric fields in their study of the VMD \opc weak chiral lagrangian. 
Instead we propose in this work to analyse the conventional
vector formulation.
\par
Let us then introduce the nonet of spin-1 resonance fields
\begin{eqnarray}
R_{\mu} \, = \, \Frac{1}{\sqrt{2}} \, \sum_{i=1}^{8} \, \lambda_i \, 
R_{\mu}^i \, + \, \Frac{1}{\sqrt{3}} \, R_{\mu}^0 \; & , & \; \; \; 
\; \; R_{\mu} \, = \, V_{\mu} , A_{\mu}~, 
\label{eq:vmu}
\end{eqnarray}
that transforms homogeneously under the chiral group as
\begin{equation}
R_{\mu} \, \mapright{G} \, h(g,\varphi) \, R_{\mu} \, 
h(g,\varphi)^{\dagger}~.
\label{eq:vmhvh}
\end{equation}
The mixing between the eighth and the singlet components of the vector
field is assumed ideal, i.e. $V_8^{\mu} \, = \, ( \, \omega^{\mu} \, +
\, \sqrt{2} \,  \phi^{\mu} \, ) \, / \, \sqrt{3}$, consistently with
the phenomenology. There is no such a statement in the axial--vector 
case but nevertheless we will do the same assumption here (in any case
deviations from this would affect very marginally our results).
\par 
The kinetic term for the resonance field is 
\begin{equation}
{\cal L}_{K} \, = \, - \, \Frac{1}{4} \, \langle \, R_{\mu \nu} \, 
R^{\mu \nu} \, \rangle \, + \, \Frac{m_R^2}{2} \, \langle \, R_{\mu} \, 
R^{\mu} \, \rangle~,
\label{eq:kterm}
\end{equation}
where $R_{\mu \nu} \, = \, \nabla_{\mu} R_{\nu} \, - \, 
\nabla_{\nu} R_{\mu}$ and $\nabla_{\mu}$ is the covariant derivative 
defined in Ref.~\cite{EG89} as
\begin{eqnarray}
\nabla_{\mu} \, A \, & = & \, \partial_{\mu} \, A \, + \, 
\left[ \Gamma_{\mu} \, , \, A \, \right]~, \nonumber \\
& & \label{eq:cova} \\
\Gamma_{\mu} \, & \equiv & \, \Frac{1}{2} \, 
\left\{ \, u^{\dagger} \, ( \partial_{\mu} \, - \, i \, r_{\mu} \, ) \, u 
\, + \, u \, ( \partial_{\mu} \, - \, i \, \ell_{\mu} \, ) \, u^{\dagger}
\, \right\}~, \nonumber
\end{eqnarray}
for any $A$ operator that transforms homogeneously as the resonant field in 
Eq.~(\ref{eq:vmhvh}).
\par
The most general strong lagrangian linear in the vector field and up to  
${\cal O}(p^3)$, assuming
nonet symmetry, reads \cite{EGL89,PradesZ}
\begin{eqnarray}
{\cal L}_V \, & = & \, - \, \Frac{f_V}{2 \sqrt{2}} \la \, V_{\mu \nu} \, 
f_{+}^{\mu \nu} \, \ra \, - \, \Frac{i g_V}{2 \sqrt{2}} \, \la 
V_{\mu \nu} \, [ \, u^{\mu} \, , \, u^{\nu} \, ] \ra  \nonumber \\
& &  \, + \, i \alpha_V \, \la \, V_{\mu} \, [ \, u_{\nu} \, , \, 
f_{-}^{\mu \nu} \, ] \, \ra 
 \, + \, \beta_V \, \la \, V_{\mu} \, 
[ \, u^{\mu} \, , \, \chi_{-} \, ] \, \ra \, \nonumber \\
& & \,  + \, h_V \, \varepsilon_{\mu \nu \rho \sigma} \, 
\la \, V^{\mu} \, \{ \, u^{\nu} \, , \, f_{+}^{\rho \sigma} \, \} \, \ra \,
+ \, i \theta_V \, \varepsilon_{\mu \nu \rho \sigma} \, \la \, 
V^{\mu} \, u^{\nu} \, u^{\rho} \, u^{\sigma} \, \ra~, 
\label{eq:evf}
\end{eqnarray}
where
\begin{equation}
f_{\pm}^{\mu \nu} \, = \, u \, F_L^{\mu \nu} \, u^{\dagger} \, \pm \, 
                        u^{\dagger} \, F_R^{\mu \nu} \, u~,
\label{eq:fplus}
\end{equation}
and $F_{R,L}^{\mu \nu}$ are the strength field tensors associated to the
external $r_{\mu}$ and $\ell_{\mu}$ fields, respectively. The couplings
in ${\cal L}_V$ can be determined, in principle, from the phenomenology
of the vector meson decays. Thus $|f_V|$, $|h_V|$, $|\theta_V|$ and 
$|\alpha_V|$ could be obtained from the experimental widths \cite{PDG}
of $\rho^0 \rightarrow e^+ e^-$, $\omega \rightarrow \pi^0 \gamma$, 
$\omega \rightarrow \pi \pi \pi$ and $\rho \rightarrow \pi \pi \gamma$,
respectively, while $g_V$ and $\beta_V$ enter in $\rho \rightarrow 
\pi \pi$. In Table 1 we collect the experimental determinations (when
available) and also the predictions of the Hidden Symmetry model (HS)
\cite{BK88} \footnote{In Ref.~\cite{DPN97} we evaluated $h_V$ in the
HS model.} and Extended Nambu--Jona--Lasinio model (ENJL) \cite{PradesZ}
\footnote{We thank F.J. Botella for pointing out to us a few mistakes
in the evaluation of the couplings in Ref.~\cite{PradesZ} and for 
providing us with his own evaluation of the couplings in the ENJL 
model.}. The relative signs of the couplings are well fixed from the
phenomenology. The positive slope of the $\pi^0 \rightarrow \gamma 
\gamma^*$ form factor implies that $f_V h_V > 0$, while the charge 
radius of the pion together with the assumption of VMD gives $f_V g_V >0$.
Finally we get $g_V \beta_V < 0$ from a combined analysis of 
$\Gamma \, (\rho \rightarrow \pi \pi)$ and $\Gamma \, (\phi 
\rightarrow K^+ K^-)$. We see that the model predictions agree with these
results. In any case we are more interested in the general 
consequences of vector meson exchange than in an accurate numerical estimate,
unrealistic with the present knowledge.

\begin{table}
\begin{center}
\begin{tabular}{|c||c|c|c|}
\hline
& HS & ENJL & Expt. \\
\hline
\hline
$f_V$ & input & $\; $0.17 & $\; $ 0.20 \\
\hline
$g_V$ & $f_V/2$ & $\; $0.08 & $\; $ 0.09 \\
\hline
$h_V$ & 0.039 & $\; \; \, $0.033 & $\; \; \,$ 0.037 \\
\hline
$\theta_V$ & $2 h_V$ & $\; \; \,$0.050 & - \\
\hline
$\alpha_V$ & - & $-$0.015 & - \\
\hline
$\beta_V$ & - & $-$0.015 & $-$0.018 \\
\hline
\end{tabular}
\caption{${\cal O}(p^3)$ vector couplings in the Hidden Symmetry model
(HS), Extended Nambu--Jona--Lasinio model (ENJL) and the experiment 
(Expt.) when available. We do not quote the experimental errors 
(typically $\simeq 10 \%$) since
higher chiral order corrections, difficult to estimate and in principle
larger, have not been considered yet.}
\end{center}
\end{table}

Analogously the most general strong lagrangian linear in the axial--vector
field and up to ${\cal O}(p^3)$, assuming
nonet symmetry, reads \cite{EGL89,PradesZ}
\begin{eqnarray}
{\cal L}_A \, & = & \, - \, \Frac{f_A}{2 \sqrt{2}} \, \la \, A_{\mu \nu} \, 
f_{-}^{\mu \nu} \, \ra \, + \, i \alpha_A \, \la \, A_{\mu} \, [ \, 
u_{\nu} \, , \, f_{+}^{\mu \nu} \, ] \, \ra 
 \, + \, \gamma_1 \, \la \, A_{\mu} \, u_{\nu} \, u^{\mu} \, u^{\nu} \, 
\ra \, \nonumber \\
& & \, + \, \gamma_2 \, \la \, A_{\mu} \, \{ \, u^{\mu} \, , \, u^{\nu} \, 
u_{\nu} \, \} \, \ra 
\, + \, \gamma_3 \, \la \, A_{\mu} \, u_{\nu} \, \ra \, \la \, 
u^{\mu} \, u^{\nu} \, \ra \, + \, \gamma_4 \, \la \, A_{\mu} \, u^{\mu} \, 
\ra \, \la \, u_{\nu} \, u^{\nu} \, \ra \, \nonumber \\
& & \, + \, h_A \, \varepsilon_{\mu \nu \rho \sigma} \, \la \, A^{\mu} \, 
\{ \, u^{\nu} \, , \, f_{-}^{\rho \sigma} \, \} \, \ra~.
\label{eq:eaf}
\end{eqnarray}

\begin{table}
\begin{center}
\begin{tabular}{|c||c|c|}
\hline
& ENJL & Expt. \\
\hline
\hline
$f_A$ & $\; \; \, $0.085 &  0.097  \\
\hline
$h_A$ & $\; \; \, $0.014 & -  \\
\hline
$\alpha_A$ & $-$0.009 & - \\
\hline
$\gamma_1$ & $\; \; \, $0.004 &  - \\
\hline
$\gamma_2$ & $-$0.010 &  - \\
\hline
$\gamma_{3,4}$ & ${\cal O}(1/\sqrt{N_c})$ & - \\
\hline
\end{tabular}
\caption{${\cal O}(p^3)$ axial--vector couplings in the Extended 
Nambu--Jona--Lasinio model (ENJL) and the experimental value (Expt.)
when available. About the experimental error of $f_A$ see the 
explanation in the caption of Table 1.}
\end{center}
\end{table}

The couplings of ${\cal L}_A$ can be obtained from the study of the
axial--vector decays.  Thus $f_A$ and $\alpha_A$ enter in $a_1 \rightarrow
\pi \gamma$, $h_A$ in $a_1 \rightarrow \pi \pi \gamma$ and 
$\gamma_i$, $i=1,...4$ in $a_1 \rightarrow \pi (\pi \pi)_{S-wave}$ and
$f_1(1285) \rightarrow \eta \pi \pi, \, K \overline{K} \pi, \, \pi 
\pi \pi \pi$.
However the experimental situation is still poor \cite{PDG} and the only
rather well known coupling is $f_A$ \cite{BBR93}. Therefore we will rely,
for the rest, in the predictions of the ENJL model \cite{PradesZ}. We
collect the ${\cal O}(p^3)$ axial--vector couplings in Table~2.
\par
We would like to emphasize the fact that the strong sector in the 
conventional vector formulation, at leading ${\cal O}(p^3)$, is much
richer than the leading ${\cal O}(p^2)$ strong action in the antisymmetric
fields \cite{EG89}. Terms analogous to $\alpha_{V,A}$, $\beta_V$, 
$\gamma_i$, $h_{V,A}$ and $\theta_V$ are not present in this last formulation
at leading order.
\par
For a further extensive and thorough exposition on $\chi$PT see
Refs.~\cite{CGE95,TONI95}.

\subsection{Non--leptonic weak interactions in $\chi$PT}
\hspace*{0.5cm}
The complete  non-leptonic $\Delta S =1$ effective weak hamiltonian
at low energies ($E \ll M_W$)  for the lightest degrees of freedom (u,d,s)
is constructed through an OPE expansion. It is given by 
a sum of products of  Wilson coefficients
$C_i(\mu)$  and  four--quark local operators $Q_i(\mu)$ $\, i=1,...6$,
\cite{revheff,CFM95,BUBU95} and reads
\begin{equation}
{\cal H}_{NL}^{|\Delta S| = 1} \; = \; 
- \, \Frac{G_F}{\sqrt{2}} \, V_{ud} V_{us}^* \, 
\sum_{i=1}^6 \, C_i(\mu) \, Q_i \; + \; h.c. \; \; \; \; . 
\label{eq:heffd}
\end{equation}
Here $G_F$ is the Fermi constant and $V_{ij}$ are elements of the 
CKM-matrix. These operators (or combinations of them) belong to the
$(8_L, 1_R)$ and $(27_L, 1_R)$ representations of the chiral group G.
The first ones give only $\Delta I = 1/2$ transitions while the seconds
generate both $\Delta I = 1/2, 3/2$ transitions. In the following we 
will neglect the contributions coming from $(27_L,1_R)$ operators.
\par
At ${\cal O}(p^2)$ in $\chi$PT, we can construct only one relevant 
$|\Delta S| =1$
operator which transforms as an octet under the chiral group
\begin{eqnarray}
{\cal L}_2^{|\Delta S|=1} \, & = & \, 4 \, G_8 \, \langle \, 
\lambda_6 \, L_1^{\mu} \, L^1_{\mu} \, \rangle \; \nonumber \\
& = & \, G_8 \, F^4 \, \langle \, \Delta \, u_{\mu} \, u^{\mu} \, 
\rangle \; , 
\label{eq:weakla} 
\end{eqnarray}
where 
\begin{equation} 
L_{\mu}^1 \, = \, \Frac{\delta S_2^{\chi}}{\delta \ell^{\mu}} \; = \; 
- \, i \, \Frac{F^2}{2} \, U^{\dagger} \, D_{\mu} \, U \; = \;  
- \, \Frac{F^2}{2} u^{\dagger} \, u_{\mu} \, u \; , 
\label{eq:leftcu}
\end{equation}
is the left--handed current associated to the ${\cal O}(p^2)$ strong 
lagrangian in Eq.~(\ref{eq:str2}) and  $\Delta \, = \, u \, \lambda_6 
\, u^{\dagger}$. From the experimental width of $K \rightarrow \pi \pi$
one gets \footnote{If \opc corrections are taken
into account, a phenomenological value of $|G_8| \simeq 6.5 \times 
10^{-6} \, \mbox{GeV}^{-2}$ is obtained \cite{KMW91}.} 
\begin{equation}
|G_8|_{K \rightarrow \pi \pi} \, \equiv |G_8| \, \simeq \, 9.2 \, 
\times \, 10^{-6} \, \mbox{GeV}^{-2} \; \; \; . 
\label{eq:g8fr}
\end{equation}
At \opc the chiral weak lagrangian has been studied in 
Refs.~\cite{KM90,EK93,Esposito90} giving
37 chiral operators $W_i$ only in the octet part
\begin{equation}
{\cal L}_4^{|\Delta S|=1} \, = \, G_8 \, F^2 \, 
\sum_{i=1}^{37} \, N_i \, W_i \; , 
\label{eq:weak4}
\end{equation}
that introduce 37 new unknown coupling constants $N_i$. Their phenomenological
determination is then very difficult. Nevertheless the theory is
already predictive at this level due to the fact that the symmetry 
provides relations between processes. Thus, for instance, the same \opc 
weak counterterm 
combination appears in the structure dependent electric amplitudes to 
$K_S\rightarrow\pi^+\pi^-\gamma$, $K^+\rightarrow\pi^+\pi^0\gamma$
and $K_L (K^\pm)\rightarrow\pi\pi \pi\gamma$
\cite{DE97,DI95,DMS93,EN92}. Unfortunately the actual status of the
phenomenology is not good enough to provide absolute quantitative 
predictions and to do so one has to rely on models.
\par
Our purpose in this paper is to study the spin--1 resonance exchange
contributions to the $N_i$ couplings and to clarify how much can be
said in a model--independent way.

\subsection{The Factorization Model and the Spin-1 resonance exchange 
contributions in the antisymmetric formulation}
\hspace*{0.5cm}
The Factorization Model (FM) has been used in the context of vector
and axial--vector meson dominance \cite{EK93}.
If we neglect the penguin contributions, justified by the $1/N_c$
expansion \cite{PI91},
we can write the octet dominant piece in Eq.~(\ref{eq:heffd}) as 
\begin{equation}
{\cal H}_{NL}^{|\Delta S|=1} \, = \, - \, 
\Frac{G_F}{2 \sqrt{2}} \, V_{ud} V_{us}^* \,  
 \, C_{-}(\mu) \, Q_{-} 
\; + \; h.c. \; ,
\label{eq:hqminus}
\end{equation}
with
\begin{equation}
Q_{-} \, = \, 4 \, (\, \overline{s}_L \, \gamma^{\mu} \, u_L \, ) 
(\, \overline{u}_L \, \gamma_{\mu} \, d_L \, ) \, - \,
 4 \, (\, \overline{s}_L \, \gamma^{\mu} \, d_L \, ) 
(\, \overline{u}_L \, \gamma_{\mu} \, u_L \, ) \; ,
\label{eq:qplusm}
\end{equation}
and $(\overline{s}_L \gamma_{\mu} u_L) \equiv \Frac{1}{2} 
\overline{s}^{\alpha} \gamma_{\mu} (1 - \gamma_5) u_{\alpha}$ ($\alpha$ is
a colour index). In a chiral gauge theory the quark bilinears in the 
$Q_{-}$ operator are given by the associated left--handed current 
\begin{equation}
\Frac{\delta S}{\delta \ell^{\mu}} \; \equiv \; 
L_{\mu} \; = \; L_{\mu}^1 \, + \, L_{\mu}^3 \, + \, L_{\mu}^5 \, + ... ,
\label{eq:currsac}
\end{equation}
(the first term $L_{\mu}^1$ was already given in Eq.~(\ref{eq:leftcu}))
where $S[U, \ell, r, s, p]$ is the low--energy strong effective action
of QCD in terms of the Goldstone bosons realization $U$ and the external
fields $\ell, r, s,p$. 
\par
Factorization, in the context of $\chi$PT, amounts to the assumption that
the product of quark bilinears in $Q_{-}$ factorizes in the
{\em current $\times$ current} form \cite{PI91} as
\begin{equation}
{\cal L}_{FM} \; = \; 4 \, k_F \, G_8 \, \langle \, \lambda \, 
\Frac{\delta S}{\delta \ell_{\mu}} \, \Frac{\delta S}{\delta \ell^{\mu}} \, 
\rangle \; + \; h.c. \; , 
\label{eq:fmgenera}
\end{equation}
where $\lambda \, \equiv \, \frac{1}{2} \, (\lambda_6 \, - \, i \lambda_7)$
and $G_8$ has been defined in Eqs.~(\ref{eq:weakla},\ref{eq:g8fr}).
The FM  in (\ref{eq:fmgenera}) gives a full prediction but for an
overall fudge factor
$k_F$ that is not given by the model. In general $k_F \simeq {\cal O}(1)$ 
and naive factorization would imply $k_F \simeq 1$. Hence 
the long time standing problem of the $\Delta  I =1/2$ enhancement
can be formulated like this: while the perturbative evaluation of 
$C_{-}(\mu)$ in  Eq.~(\ref{eq:hqminus}) implies a
small value of $k_F$ \cite{CFM95,BUBU95}
\begin{equation}
C_{-}(m_\rho) \, \simeq \, 2.2 \; \; \; \; \; \; \longrightarrow 
\; \; \; \; \; \; k_F \, \simeq \, 0.2-0.3~,
\label{eq:wilson}
\end{equation}
the phenomenology of the $K\rightarrow \pi\pi(\pi)$ processes requires  
$k_F \simeq 1$. In this case is rather clear that the FM only provides
a parameterization without any physical insight in the problem, i.e. 
parameterizes uncontrolled non--factorizable contributions. However
if in other processes one finds instead that $k_F \simeq 0.2-0.3$ (as 
predicted by the Wilson coefficient) then factorization has a clear 
dynamical meaning.
\par
The r\^ole of factorization is also important in uncovering the physical
significance of the chiral structures of the weak lagrangian. For instance
the operators $W_{28}, ...W_{31}$ in ${\cal L}_4^{|\Delta S|=1}$ have
been proven to have an interesting dynamical origin~: they can be generated
by the chiral anomaly \cite{CH90,ENP94} as a factorizable contribution
in which one of the currents comes from the Wess--Zumino--Witten 
action \cite{WZW71}. Moreover in this particular case non--factorizable
contributions cannot give different chiral structures to the operators.
\par
In Ref.~\cite{EK93} the resonance exchange contributions to 
${\cal L}_4^{|\Delta S|=1}$ have been studied using the antisymmetric
formulation for the spin--1 fields. After a model--independent analysis
that specifies the $N_i$ couplings in terms of a rather large set
of unknown weak couplings involving resonances (that is already able to 
provide a list of relations between the couplings) and in order to increase 
the predictive power, the authors of that reference use the FM to be
able to specify the $N_i$ couplings in terms of only one parameter~: the
$k_F$ factorization factor in Eq.~(\ref{eq:fmgenera}).
\par 
As already pointed out in Ref.~\cite{EK93}, there are two ways to derive a  
FM weak lagrangian generated by resonance exchange (we focus here on the
spin--1 fields)~:
\begin{itemize}
\item[(${\cal A}$)] To evaluate the  strong action generated by resonance 
exchange,
and then perform the factorization procedure in Eq.~(\ref{eq:fmgenera}).
Since we apply the FM procedure once the vectors have already been 
integrated out, we can say that we have performed
factorization at the scale of the  low energy effective field theory i.e.
the Goldstone boson mass scale. 
Consequently the lagrangian has been generated at the kaon mass scale.
\item[(${\cal B}$)] Conversely we can write down the spin--1 strong chiral 
lagrangian, we apply factorization and thus we derive the weak resonance 
couplings. Then we integrate out the resonance fields and generate the 
effective lagrangian that, in this way, is generated at the scale of the 
resonance.
\end{itemize}

In principle the two effective actions, thus generated, do not have to 
coincide.
\par
Actually we have shown in a previous work \cite{DPN97} that 
the weak \ops  VMD lagrangian for  $K_L\rightarrow \gamma \gamma^*$
and $K_L\rightarrow \pi^0 \gamma \gamma $
obtained with procedure \prb has extra chiral structures 
compared to the ones obtained with procedure \pra. Both give a good 
phenomenological description,
but with different factorization factor: $k_F \simeq 1$ for method \pra,
while $k_F \simeq 0.2-0.3$  (i.e. the Wilson coefficient) for the 
scheme \prb. Also the  weak \ops  VMD lagrangian for
$K_L\rightarrow \pi^+ \pi^- \gamma$ shares this property:
procedure \prb has a more complete set of operators \cite{DPN98}
compared to the one obtained with procedure \pra \cite{ENP94}. 
Method \prb is what we define as factorization in the vectors: FMV model
\cite{DPN97}.
\par
To generate ${\cal L}_4^{|\Delta S|=1}$ in Eq.~(\ref{eq:weak4}) procedure 
\pra requires the determination of the VMD strong \opc action, i.e.
the VMD contribution to the ${\cal L}_4$ in Eq.~(\ref{eq:strong4}).
This VMD action has already been determined in the antisymmetric formulation
\cite{EG89}. If resonance saturation of ${\cal L}_4$ is assumed, both 
procedures \pra and \prb generate the same ${\cal L}_4^{|\Delta S|=1}$ 
effective action in this formulation \cite{EK93}. 
However a word of caution is necessary in relation with
the $H_1$ coupling in ${\cal L}_4$. This coupling also receives a resonant
contribution \cite{EG89} but, as already commented before, is not 
measurable. The authors of Ref.~\cite{EK93} suggested to leave free this
coupling (as we will comment below, in their phenomenological study of
$K \rightarrow \pi \gamma^*$ they input $H_1 = 0$ at the scale $\mu = 
m_{\rho}$). This amounts to add local contributions into the 
lagrangian (therefore spoiling the assumption of resonance saturation)
and then \pra and \prb schemes differ in this case.
\par
It should also be mentioned that in Ref.~\cite{EK93} the scalar and 
pseudoscalar resonance exchange were also studied. When potentially
relevant ($K\rightarrow \pi \pi (\pi)$), we will comment on it.

\section{\opc weak counterterms from spin--1 resonances in the vector
formulation}
\hspace*{0.5cm}
The ${\cal O}( p^3 )$ strong lagrangian of vectors and axial--vectors 
has been given in Eqs.~(\ref{eq:evf},\ref{eq:eaf}) respectively. Thus 
to describe vector and axial--vector exchange contributions at \opc in the
weak chiral lagrangian ${\cal L}_{4}^{|\Delta S|=1} $ in Eq.~(\ref{eq:weak4})
we need  an ${\cal O}( p )$ CP conserving effective weak lagrangian
linear in the vector and axial--vector fields.
\par
Indeed, assuming octet dominance, there are only two such operators
transforming as $(8_L,1_R)$ under the chiral group~:
\begin{equation}
{\cal L}_R^{{\cal O}(p)} \, = \, G_8 \, F_{\pi}^4 \, 
\left[ \, \omega_1^R \, \langle \, \Delta \, \{ \, R_{\mu} \, , \, 
u^{\mu} \, \} \, \rangle \, + \, \omega_2^R \, \langle \, \Delta
\, u_{\mu} \, \rangle \langle \,  R^{\mu} \, \rangle \, \right]~,
\label{eq:vpp1}
\end{equation}
where $R_{\mu}= V_{\mu}, A_{\mu}$, and
$\Delta$ has been defined in connection with Eq.~(\ref{eq:weakla}).
The operators in Eq.~(\ref{eq:vpp1}) generate ${\cal O}( p )$ weak decays of 
vectors (axials) into two or more
pseudoscalars with unknown couplings $\omega_1^R$ and $\omega_2^R$. 
To evaluate the vector (axial) contribution to 
\opc weak lagrangian ${\cal L}_{4}^{|\Delta S|=1} $ (\ref{eq:weak4}) we can
just integrate out the resonance fields between ${\cal L}_R^{{\cal O}(p)}$
and ${\cal L}_V$ and ${\cal L}_A$ in Eqs.~(\ref{eq:evf},\ref{eq:eaf})
respectively. Equivalently we can derive the \opc weak lagrangian just 
redefining the vector (axial) field as
\begin{equation}
R_{\mu } \, \longrightarrow \, R_{\mu} \, - \, 
\Frac{G_8 F_{\pi}^4}{m_R^2} \, \left[ \, \omega_1^R \, \{ \, \Delta \, ,
\, u_{\mu} \, \} \, + \, \omega_2^R \, \langle \, \Delta \, u_{\mu} \, 
\rangle \, \right]~,
\label{eq:shiftv}
\end{equation}
in ${\cal L}_V$ and ${\cal L}_A$.
\par
The relevant features of this procedure are the following~:
\begin{itemize}
\item[i)] The weak field redefinitions eliminate the weak ${\cal O}(p)$
resonance couplings in ${\cal L}_R^{{\cal O}(p)}$ Eq.~(\ref{eq:vpp1})
generating vector and axial--vector exchange contributions to the 
\opc weak lagrangian ${\cal L}_4^{|\Delta S|=1}$.
\item[ii)] The kinetic term  of the resonances in Eq.~(\ref{eq:kterm}) 
generates also an ${\cal O}( p^3 )$ weak lagrangian for the vectors (axials)
irrelevant at \opc, but necessary to derive in a complete and consistent manner
the  weak \ops  VMD lagrangian, as we will comment at the end of this Section. 
\end{itemize}

In this formulation the non-vanishing contributions to 
${\cal L}_{4}^{|\Delta S|=1}$ in Eq.~(\ref{eq:weak4})
satisfy the following relations:

\begin{itemize}
\item[1/] Vector contributions.
\begin{eqnarray}
N_1^V \, & = & - \, N_2^V \, = \, - \, N_{19}^V \, = \, N_{16}^V \, + \, 
\Frac{N_{17}^V}{2}~, \nonumber \\
2 N_{14}^V \, & = & \, N_{15}^V \, = \, - 4 \, N_{18}^V \, = \, N_{25}^V \, 
= \, 4 \, N_{27}^V \,= \,- 4 \, N_{37}^V~, \nonumber  \\
 N_{29}^V \, & = & \, N_{34}^V~, 
\label{eq:gva} \\ 
g_V \, N_{17}^V \, & = &-2\, \sqrt{2} \alpha_V \, N_{1}^V~, \nonumber \\
f_V\, N_{29}^V \, & = & \, \frac{h_V}{\sqrt{2}}\, N_{15}^V \; ~,
\nonumber \\
f_V\, N_{1}^V \, & = & \, -g_V\, N_{15}^V \; ~,
\nonumber \\
2\, h_V \, N_{28}^V \, & = & \, \theta_V \, N_{30}^V  \; ~,
\nonumber \\
g_V \, N_9^V \, & = & \, - \, \sqrt{2} \, \beta_V \, N_1^V \; ~. 
\nonumber
\end{eqnarray}
\item[2/] Axial--vector contributions.
\begin{eqnarray}
N_{15}^A \, & = &  - 2 \, N_{14}^A~, \nonumber \\
2 N_{16}^A \, & = & N_{17}^A \, = \, \Frac{4}{3} \, N_{18}^A \, = \, 
N_{26}^A \, = \, - 4 \, N_{27}^A \, = \, -4 \, N_{37}^A~,
\nonumber\\
N_{29}^A \, & = & - N_{34}^A~,  
\label{eq:axials} \\
2 \sqrt{2}\, \alpha_A \,N_{17}^A \, & = &  \, f_A \, N_{15}^A \, \nonumber \\
 h_A \,N_{17}^A \, & = & - \, \sqrt{2} \, f_A \, N_{29}^A~. \nonumber 
\end{eqnarray}
\end{itemize}
Comparing these relations with the ones obtained by Ecker et al. in the
antisymmetric formulation \cite{EK93} (see their Eqs.~(3.18,3.19) for the
vectors and Eq.~(3.20) for the axials) we see that most of their relations
are satisfied in our framework using $f_V = 2 g_V$ (as they do). The only
exceptions correspond to our results in the combinations
$N_1^V - 2 N_2^V + 3 N_{14}^V + 3 N_{16}^V - 6 N_{18}^V = - 3 F_{\pi}^2 \,
\alpha_V \, \omega_1^V / m_V^2$ and $N_{16}^V - N_{18}^V + N_{27}^V = 
- F_{\pi}^2 \, \alpha_V \, \omega_1^V / m_V^2$ both of which they find to
vanish. In any case our relations in Eqs.~(\ref{eq:gva},\ref{eq:axials})
are many more (some of them independent of the assumptions $f_V = 2 g_V$
and $\theta_V = 2 h_V$). This is due to the fact that while we have
a richer structure on the strong sector (as emphasized before) our leading
weak ${\cal O}(p)$ lagrangian ${\cal L}_R^{{\cal O}(p)}$ in 
Eq.~(\ref{eq:vpp1}) is much more restricted in comparison with the leading
${\cal O}(p^2)$ weak lagrangian for spin--1 resonance contribution
in the antisymmetric formalism \cite{EK93}. Both realizations get therefore
a balance. The vector formalism is richer in the strong sector while
the antisymmetric tensor fields give a richer structure to the weak one.
The fact that the main r\^ole is given to the rather well known strong
couplings in the conventional vector fields offers a much more predictive
scheme in a model--independent way.
\par
In the third and fourth columns of Tables 3,4 and 5  we report the spin--1
resonance contributions to the couplings $N_i$ of the phenomenologically
relevant $W_i$ operators. For the ones that are not involved in the 
relevant decays~: $N_{19}$, $N_{25}$, $N_{26}$, $N_{27}$, $N_{34}$ and
$N_{37}$ we refer to the Eqs.~(\ref{eq:gva},\ref{eq:axials}) above (the
structures of the corresponding $W_i$ operators can be read from 
Ref.~\cite{EK93}). The remaining operators in ${\cal L}_4^{|\Delta S|=1}$
do not get any contributions.

\begin{sidewaystable}
\begin{center}
\begin{tabular}{|c|c|c|c|c|c|} 
\hline
\multicolumn{1}{|c|}{} & 
\multicolumn{1}{|c|}{} & 
\multicolumn{1}{|c|}{} & 
\multicolumn{1}{|c|}{} & 
\multicolumn{2}{|c|}{Expressions using} \\  
\multicolumn{1}{|c|}{$i$} & 
\multicolumn{1}{|c|}{$W_i$} & 
\multicolumn{1}{|c|}{Vectors} & 
\multicolumn{1}{|c|}{Axial--Vectors} & 
\multicolumn{2}{|c|}{$\omega_1^R = \sqrt{2} \Frac{m_R^2}{F_{\pi}^2}
f_R \eta_R$, $\; \; $  $\omega_2^R = - \omega_1^R$} \\
\cline{5-6}
\multicolumn{1}{|c|}{} & 
\multicolumn{1}{|c|}{} & 
\multicolumn{1}{|c|}{} & 
\multicolumn{1}{|c|}{} & 
\multicolumn{1}{|c|}{Vectors} & 
\multicolumn{1}{|c|}{Axial--Vectors} \\ 
\hline
\hline
& & & & & \\
1 & $\la \Delta  u_\mu u^\mu u_\nu u^\nu \ra$ &
$- \Frac{F_{\pi}^2}{m_V^2} \,  \Frac{g_V}{\sqrt{2}} \, \omega_1^V$ & 
$\Frac{2}{3} \, \Frac{F_{\pi}^2}{m_A^2} \, (\gamma_1 - 3 \gamma_2 - 4 \gamma_4) 
\, \omega_1^A$ & $- f_V g_V \eta_V $ & $\Frac{2 \sqrt{2}}{3} f_A
(\gamma_1 - 3 \gamma_2 - 4 \gamma_4) \eta_A$ \\ 
& & & & & \\
\hline
& & & & & \\
2 & $\la \Delta u_{\mu} u_{\nu} u^{\nu} u^{\mu} \ra$ &
$\Frac{F_{\pi}^2}{m_V^2} \, \Frac{g_V}{\sqrt{2}} \, \omega_1^V$ &
$\Frac{2}{3} \,  \Frac{F_{\pi}^2}{m_A^2} \, ( 2\gamma_1 - 3 \gamma_2 - 
2 \gamma_4) \, 
\omega_1^A$ & $ f_V g_V \eta_V$ & $\Frac{2 \sqrt{2}}{3} f_A
(2 \gamma_1 - 3 \gamma_2 - 2 \gamma_4) \eta_A$ \\
& & & & & \\
\hline
& & & & & \\
3 & $\la \Delta u_{\mu} u_{\nu} \ra \la u^{\mu} u^{\nu} \ra$ &
 -  & $-2 \Frac{F_{\pi}^2}{m_A^2} \, (\gamma_1 + \gamma_3) \, \omega_1^A$ &
 -  & $- 2 \sqrt{2} f_A (\gamma_1 + \gamma_3) \eta_A$ \\
& & & & & \\
\hline
& & & & & \\
4 & $\la \Delta u_{\mu} \ra \la u^{\mu} u_{\nu} u^{\nu} \ra$ &
 -  & $ - \Frac{F_{\pi}^2}{m_A^2} \left[ \Frac{4}{3} (\gamma_1 - \gamma_4)
\, \omega_1^A + (\gamma_1 + 2 \gamma_2) \, \omega_2^A \right] $ & - & 
$- \Frac{\sqrt{2}}{3} f_A (\gamma_1 - 6 \gamma_2 - 4 \gamma_4) \eta_A$ \\
& & & & & \\
\hline
& & & & & \\
9 & $\la \Delta [ \chi_{-}, u_{\mu} u^{\mu} ] \ra$ & 
$\Frac{F_{\pi}^2}{m_V^2} \, \beta_V \, \omega_1^V$ & - & 
$\sqrt{2} f_V \beta_V \eta_V $ & - \\
& & & & & \\
\hline
\end{tabular} 
\caption{Vector and Axial--vector contribution to the $N_i$ coefficients of 
the $W_i$ octet operators, in the 
basis of Ref.~\protect\cite{EK93}, relevant to pure
non--leptonic kaon decays at $O(G_F)$. The hypothesis of factorization 
is only used to relate $\omega_1^A$ with $\omega_2^A$ in the operator 
$W_4$. }
\end{center}
\end{sidewaystable}

\begin{sidewaystable}
\begin{center}
\begin{tabular}{|c|c|c|c|c|c|} 
\hline
\multicolumn{1}{|c|}{} & 
\multicolumn{1}{|c|}{} & 
\multicolumn{1}{|c|}{} & 
\multicolumn{1}{|c|}{} & 
\multicolumn{2}{|c|}{Expressions using} \\  
\multicolumn{1}{|c|}{$i$} & 
\multicolumn{1}{|c|}{$W_i$} & 
\multicolumn{1}{|c|}{Vectors} & 
\multicolumn{1}{|c|}{Axial--Vectors} & 
\multicolumn{2}{|c|}{$\omega_1^R = \sqrt{2} \Frac{m_R^2}{F_{\pi}^2}
f_R \eta_R$} \\
\cline{5-6}
\multicolumn{1}{|c|}{} & 
\multicolumn{1}{|c|}{} & 
\multicolumn{1}{|c|}{} & 
\multicolumn{1}{|c|}{} & 
\multicolumn{1}{|c|}{Vectors} & 
\multicolumn{1}{|c|}{Axial--Vectors} \\ 
\hline
\hline
& & & & & \\
14 & $i \la \Delta \{ f_{+}^{\mu \nu} , u_{\mu} u_{\nu} \} \ra$ &
$ \Frac{F_{\pi}^2}{m_V^2} \, \Frac{f_V}{2 \sqrt{2}} \, \omega_1^V$ & 
$ - \Frac{F_{\pi}^2}{m_A^2} \, \alpha_A \, \omega_1^A$ & 
$ \Frac{1}{2} f_V^2 \eta_V$ & 
$ - \sqrt{2} f_A \alpha_A \eta_A$ \\
& & & & & \\
\hline
& & & & & \\
15 & $i \la \Delta u_{\mu} f_{+}^{\mu \nu} u_{\nu} \ra$ &
$\Frac{F_{\pi}^2}{m_{V}^2} \, \Frac{f_V}{\sqrt{2}} \, \omega_1^V$ &
$ 2 \Frac{F_{\pi}^2}{m_A^2} \, \alpha_A \, \omega_1^A $ &
$ f_V^2 \eta_V $ & 
$ 2 \sqrt{2} f_A \alpha_A \eta_A$ \\
& & & & & \\
\hline
& & & & & \\
16 & $i \la \Delta \{ f_{-}^{\mu \nu} , u_{\mu} u_{\nu} \} \ra$ & 
$ - \Frac{F_{\pi}^2}{m_V^2} \left( \Frac{g_V}{\sqrt{2}} + \alpha_V \right)
\omega_1^V$ &
$ \Frac{F_{\pi}^2}{m_A^2} \, \Frac{f_A}{2 \sqrt{2}} \, \omega_1^A$ &
$ - f_V ( g_V + \sqrt{2} \alpha_V ) \eta_V $ &
$ \Frac{1}{2} f_A^2 \eta_A$ \\
& & & & & \\
\hline
& & & & & \\
17 & $i \la \Delta u_{\mu} f_{-}^{\mu \nu} u_{\nu} \ra$ & 
$ 2 \Frac{F_{\pi}^2}{m_V^2} \, \alpha_V \, \omega_1^V $ & 
$ \Frac{F_{\pi}^2}{m_A^2} \, \Frac{f_A}{\sqrt{2}} \, \omega_1^A$ &
$ 2 \sqrt{2} f_V \alpha_V \eta_V$ &
$ f_A^2 \eta_A $ \\
& & & & & \\
\hline
& & & & & \\
18 & $\la \Delta ( f_{+ \, \mu \nu}^2 - f_{- \, \mu \nu}^2 ) \ra$ &
$ - \Frac{F_{\pi}^2}{m_V^2} \, \Frac{f_V}{4 \sqrt{2}} \, \omega_1^V$ &
$ \Frac{3}{4 \sqrt{2}} \, \Frac{F_{\pi}^2}{m_A^2} \, f_A \, \omega_1^A$ & 
$ - \Frac{1}{4} f_V^2 \eta_V $ &
$ \Frac{3}{4} f_A^2 \eta_A$ \\
& & & & & \\
\hline
\end{tabular} 
\caption{Vector and Axial--vector contribution to the $N_i$ coefficients of the
$W_i$ octet operators, in the 
basis of Ref.\protect\cite{EK93}, relevant to radiative
non--anomalous non--leptonic kaon decays at $O(G_F)$. Notice that the
factorization hypothesis is not used anywhere in this Table.}
\end{center}
\end{sidewaystable}

\begin{sidewaystable}
\begin{center}
\begin{tabular}{|c|c|c|c|c|c|} 
\hline
\multicolumn{1}{|c|}{} & 
\multicolumn{1}{|c|}{} & 
\multicolumn{1}{|c|}{} & 
\multicolumn{1}{|c|}{} & 
\multicolumn{2}{|c|}{Expressions using} \\  
\multicolumn{1}{|c|}{$i$} & 
\multicolumn{1}{|c|}{$W_i$} & 
\multicolumn{1}{|c|}{Vectors} & 
\multicolumn{1}{|c|}{Axial--Vectors} & 
\multicolumn{2}{|c|}{$\omega_1^R = \sqrt{2} \Frac{m_R^2}{F_{\pi}^2}
f_R \eta_R$, $\; \; $  $\omega_2^R = - \omega_1^R$} \\
\cline{5-6}
\multicolumn{1}{|c|}{} & 
\multicolumn{1}{|c|}{} & 
\multicolumn{1}{|c|}{} & 
\multicolumn{1}{|c|}{} & 
\multicolumn{1}{|c|}{Vectors} & 
\multicolumn{1}{|c|}{Axial--Vectors} \\ 
\hline
\hline
& & & & & \\
28 & $ i \varepsilon_{\mu \nu \rho \sigma} \la \Delta u^{\mu} \ra
\la u^{\nu} u^{\rho} u^{\sigma} \ra$ & 
$ - \Frac{F_{\pi}^2}{m_V^2} \, \theta_V \, \omega_2^V $ &
-  & $ \sqrt{2} \, f_V \theta_V \eta_V$ & -  \\
& & & & & \\
\hline
& & & & & \\
29 & $ \varepsilon_{\mu \nu \rho \sigma} \la \Delta \left[ f_{+}^{\rho 
\sigma} - f_{-}^{\rho \sigma} , u^{\mu} u^{\nu} \right] \ra$ &
$ \Frac{F_{\pi}^2}{m_V^2} \, \Frac{h_V}{2} \, \omega_1^V$ & 
$ - \Frac{F_{\pi}^2}{m_A^2} \, \Frac{h_A}{2} \, \omega_1^A$ &
$ \Frac{1}{\sqrt{2}} \, f_V h_V \eta_V$ & 
$ - \Frac{1}{\sqrt{2}} \,  f_A h_A \eta_A$ \\
& & & & & \\
\hline
& & & & & \\
30 & $ \varepsilon_{\mu \nu \rho \sigma} \la \Delta u^{\mu} \ra \la
f_{+}^{\rho \sigma} u^{\nu} \ra$ & 
$ - 2 \Frac{F_{\pi}^2}{m_V^2} \, h_V \, \omega_2^V$ &
-  & $ 2 \sqrt{2} \, f_V h_V \eta_V$ & -  \\
& & & & & \\
\hline
& & & & & \\
31 & $ \varepsilon_{\mu \nu \rho \sigma} \la \Delta u^{\mu} \ra \la
f_{-}^{\rho \sigma} u^{\nu} \ra$ & 
-  & $- 2 \Frac{F_{\pi}^2}{m_A^2} \, h_A \, \omega_2^A$ &  -  &
$ 2 \sqrt{2} \, f_A h_A \eta_A$ \\
& & & & & \\
\hline
\end{tabular}
\caption{Vector and Axial--vectors contribution to the $N_i$ coefficients 
of the $W_i$ octet operators, in the 
basis of Ref.\protect\cite{EK93}, relevant to radiative
anomalous non--leptonic kaon decays at $O(G_F)$.
The hypothesis of factorization 
is only used to relate $\omega_1^R$ with $\omega_2^R$. }
\end{center}
\end{sidewaystable}

In our framework the r\^ole of the model dependence (factorization) has a
very marginal impact~: one finds only a relation among $\omega_1^R$ and
$\omega_2^R$ ($R= V,A$) in ${\cal L}_R^{{\cal O}(p)}$ in 
Eq.~(\ref{eq:vpp1}). Indeed we find (see Appendix A)~:
\begin{equation}
\omega_1^R \, = \, \sqrt{2} \, \Frac{m_R^2}{F_{\pi}^2} \, f_R \, \eta_R \; 
= \, - \, \omega_2^R~, \; \; \; \; \; \; \; \quad R=V,A~,
\label{eq:omegav}
\end{equation}
where $\eta_R, R=V,A$ is the ${\cal O}(1)$ unknown factorization factor
(in principle $\eta_V \neq \eta_A$). If the $\Delta I = 1/2$ 
enhancement is at work then $\eta_R \simeq 1$, while $\eta_R \simeq
0.2-0.3$ if it is given by the Wilson coefficient. In Eq.~(\ref{eq:omegav})
$f_R$, $R=V,A$ are the strong couplings defined in ${\cal L}_V$ 
(Eq.~(\ref{eq:evf})) and ${\cal L}_A$ (Eq.~(\ref{eq:eaf})).
\par
In the fifth and sixth column of Tables 3,4 and 5 we rewrite the 
spin--1 resonance contribution to $N_i$ using Eq.~(\ref{eq:omegav}) for
the relevant phenomenological operators.
\par
Notice that in those couplings where only one of the two $\omega_i^R$
appears (independently for $R=V,A$), i.e. all the relatives to 
non--radiative non--leptonic kaon decays (Table 3) (but for the axial
contribution to $N_4$), and all the relatives to the non--anomalous
radiative non--leptonic kaon decays (Table 4) the use of factorization
(Eq.~(\ref{eq:omegav})) just amounts to rewriting the couplings
without extra information~: they are model--independent. 
\par
The assumption of factorization only enters in the
operators in Table 5 that contribute to anomalous radiative non--leptonic
kaon decays ($N_{28}^V$ and $N_{30}^V$ depend on $\omega_2^V$ and 
$N_{31}^A$ depend on $\omega_2^A$). In fact we get two extra relations~:
\begin{eqnarray}
N_{30}^V & = & 4 \, N_{29}^V~,  \nonumber \\
N_{31}^A & = & -4 \, N_{29}^A~.  
\label{eq:FMNA}
\end{eqnarray}
Also by comparison to Ref.~\cite{EK93} we do get vector and 
axial--vector contributions
to the anomalous $N_{28},..N_{31}$, while factorization in the antisymmetric
formulation gives a vanishing contribution.
In the remaining vector  contributions  we agree completely,
while for the axial--vectors  we disagree  with Ref.~\cite{EK93} only
in $N_{16}^A$ and $N_{17}^A$: $N_{17}^A$ is vanishing and $N_{16}^A$
is twice our value in their formulation. 
We stress also that for the other axial contributions like $N_{18}^A$ 
we agree, and for us the relative weights 
among all axial couplings, (see Eq.~(\ref{eq:axials})) are independent of 
factorization.
Thus we attribute this difference
to the  diverse formulation; of course this has physical consequences,
as we will comment in the discussion on the phenomenology.
\par
Regarding the two different procedures: (${\cal A}$) to integrate the 
vectors first and then  perform FM, or (${\cal B}$) vice versa (see 
discussion at the end of Section 2.3)
here, so far we have used only procedure \prb.
If we assume that the \opc  VMD strong lagrangian is the same
as the antisymmetric formulation \cite{EGL89}
procedures \prb and \pra differ just as much as we differ
from the FM results of Ref.~\cite{EK93}. 
\par
As already commented at the end of Section 2.3, in the 
weak \ops  VMD   lagrangian for  $K_L\rightarrow \gamma \gamma^*$
and $K_L\rightarrow \pi^0 \gamma \gamma $
procedure \prb produces all the structures generated by procedure \pra,
plus additional ones. In fact the  ${\cal O}( p^3 )$ vector weak lagrangian
induced by the shift in Eq.~(\ref{eq:shiftv}) in the kinetic term is 
needed in order to recover completely in procedure \prb
the structure generated by procedure \pra \cite{ENP94}, 
as we have shown explicitly in the weak \ops  VMD lagrangian for
$K_L\rightarrow \pi^+ \pi^- \gamma$ \cite{DPN98}.
\par
This gives, we think, also  more reliability to our model which 
describes simultaneously,
in a consistent and complete manner, the weak \opc and \ops  VMD lagrangians 
in the conventional vector formulation. 

\section{Discussion on the Phenomenology}
\hspace*{0.5cm}
The new features of the framework we have proposed in the last Section
provide important consequences on the phenomenology of many non--leptonic
kaon decays. We will discuss them here in turn.

\subsection{$K\rightarrow \pi \pi(\pi)$}
\hspace*{0.5cm}
$K\rightarrow  3\pi$ amplitudes, due to the small phase space available,
are generally expanded in the kinetic
energy of the pions up to quadratic slopes
and decomposed, jointly with $K\rightarrow \pi\pi $ ,
according to the isospin of the final state \cite{DAGI96,KMW91,MP95}.
Then the isospin amplitudes are determined by fitting the full set 
of data \cite{KMW91}. 
If  we neglect the $\Delta I =3/2$ transitions,
the determined  amplitudes are  $A_0$ for $K\rightarrow \pi\pi $
and in $K\rightarrow  3\pi$ the  amplitude at the center of the Dalitz
plot ($\alpha_1$), the linear slope ($\beta_1$) and the  
quadratic slopes ($\zeta_1$  and $\xi_1$). 
\par
The chiral expansion proves to be predictive.
$K\rightarrow \pi \pi(\pi)$ start at ${\cal O}$($p^2$) in $\chi$PT and
receive sizable \opc loop and counterterm contributions  \cite{KMW91}. 
The counterterms have two possible sources:
\begin{itemize}
\item[i)] Weak transitions in the external legs of pole diagrams that are
determined by the strong $L_i$ couplings in ${\cal L}_4$.
\item[ii)] Direct weak terms that are provided by the weak $N_i$ couplings
in ${\cal L}_4^{|\Delta S|=1}$.
\end{itemize}
The experimentally determined $\Delta I =1/2$ amplitudes and slopes 
receive the 
following weak \opc counterterm contributions \cite{KMW91}~:
\begin{eqnarray}
A_0 \, & = & \, - \, \Frac{m_K^2}{F_{\pi}^2 F_K} \, 
\Frac{2}{3} \sqrt{\Frac{2}{3}} \,  
( m_K^2 - m_{\pi}^2 ) \, K_1~, \nonumber \\
\alpha_1 \, & = & \, - \, \Frac{2 \, m_K^4}{27 F_K F_{\pi}^3}  \, 
\left[ \, (K_1 \, - \, K_2) \, + \, 24 \, G_8 F_{\pi}^2 \, (2 L_1 \, + \, 
2 L_2 \, + \,  L_3 ) \, \right]~, \nonumber \\
\beta_1 \, & = & \, - \, \Frac{m_K^2 m_{\pi}^2}{9 F_K F_{\pi}^3} \, 
\left[ \, (K_3 \, - \, 2 K_1) \, + \, 24 \, G_8 F_{\pi}^2 \, (-2 L_1 \, + \, 
L_2 \, - \, L_3 \, + \, 12 L_4) \, \right]~, \nonumber \\
\zeta_1 \, & = & \, - \, \Frac{m_{\pi}^4}{6 F_K F_{\pi}^3} \, 
\left[ \, K_2 \, - \, 24 \, G_8 F_{\pi}^2 \, (2 L_1 \, + \, 2 L_2 \, + 
\, L_3) \, \right]~, \nonumber \\
\xi_1 \, & = & \, - \, \Frac{m_{\pi}^4}{6 F_K F_{\pi}^3} \, 
\left[ \, K_3 \, - \, 24 \, G_8 F_{\pi}^2 \, ( 2 L_1 \, - \, L_2 \, + 
\, L_3) \, \right]~. 
\label{eq:k23pi}
\end{eqnarray}
Here $K_1, \ K_2$ and $K_3$ are three scale dependent weak \opc counterterm
combinations in terms of $N_i$. In the second column of Table 6 we show the
specific dependence, while in the last three columns we collect their
determination from the fit to experiment rates and slopes obtained by 
Ecker et al. \cite{EK93} for three different scales, i.e. $\mu = m_{\eta},
m_{\rho}, 1 \, \mbox{GeV}$.
 
\begin{sidewaystable}
\begin{center}
\begin{tabular}{|c|c|c|c|c|c|c|c|} 
\hline
\multicolumn{1}{|c|}{} & 
\multicolumn{1}{|c|}{} & 
\multicolumn{2}{|c|}{Our model--independent analysis} &  
\multicolumn{4}{|c|}{Ecker et al. analysis \cite{EK93}} \\
\cline{3-8}
\multicolumn{1}{|c|}{$i$} &
\multicolumn{1}{|c|}{$K_i$} & 
\multicolumn{2}{|c|}{V$+$A (vector formulation)} & 
\multicolumn{1}{|c|}{S$+$P} & 
\multicolumn{3}{|c|}{Phenomenological result} \\
\cline{3-8}
\multicolumn{1}{|c|}{} & 
\multicolumn{1}{|c|}{} & 
\multicolumn{1}{|c|}{Prediction} &  
\multicolumn{1}{|c|}{Numerical value} & 
\multicolumn{1}{|c|}{(FM)} & 
\multicolumn{1}{|c|}{$\mu = m_{\eta}$} & 
\multicolumn{1}{|c|}{$\mu = m_{\rho}$} & 
\multicolumn{1}{|c|}{$\mu = 1$ GeV} \\
\hline
\hline
& & & & & & &  \\
1 & $9 G_{8} F_{\pi}^2 ( -N_5^r + 2 N_7^r - 2 N_8^r - N_9^r )$ & 
$-9 \sqrt{2} G_8 F_{\pi}^2 f_V \beta_V \eta_V$ &
$3.65 \eta_V$ &
$-1.70 k_F$ & 
$0.0$ & $4.5$ & $7.9$ \\
& & & & & & & \\
\hline
& & & & & & & \\
2 & $3 G_8 F_{\pi}^2 ( N_1^r + N_2^r + 2 N_3) $ & 
$-6 \sqrt{2} G_8 F_{\pi}^2 (\gamma_1 + 2 \gamma_2 ) f_A \eta_A $ & 
$1.05 \eta_A$ & 
$1.70 k_F$ 
& $8.4$ & $7.8$ & $7.3$ \\
& & & & & & & \\
\hline
& & & & & & & \\
3 & $3 G_8 F_{\pi}^2 ( N_1^r + N_2^r - N_3 ) $ & 
$ 12 \sqrt{2} G_8 F_{\pi}^2 (\gamma_1 - \gamma_2 ) f_A \eta_A$ & 
$1.83 \eta_A$ & 
$1.70 k_F$ & 
$5.8$ & $5.2$ & $4.7$ \\
& & & & & & & \\
\hline
\end{tabular} 
\caption{Counterterm combinations $K_i$ relevant for 
$K \rightarrow \pi \pi / \pi \pi \pi$. The numerical value of our 
predictions is in units of
$[G_8/(9.2 \times 10^{-6} \, \mbox{GeV}^{-2})] \times 10^{-9}$. The
numerical values in the Ecker et al. analysis are in units of 
$10^{-9}$. Our analysis gives the vector and axial--vector 
contributions ($V+A$) in the model--independent framework proposed
in the text. In the third column we have 
put $\gamma_3 = \gamma_4 = 0$ in $K_2$ and $K_3$. The analysis of Ecker 
et al. \protect\cite{EK93} gives the 
contributions of scalar and pseudoscalar ($S+P$) resonances in the FM.}
\end{center}
\end{sidewaystable}

The first analysis of \opc weak VMD through vector resonances 
(no axial--vectors) in  $K\rightarrow \pi\pi\pi$
was performed in the hidden symmetry formulation of vector mesons 
\cite{IP92}, and the FM was then used to evaluate the couplings. 
However no vector contribution to the direct vertices has been found in 
this framework.
\par
A complete analysis of vector, axial--vector, scalar and pseudoscalar
resonances in the FM has been done in Ref.~\cite{EK93} using the
antisymmetric formulation for the spin--1 fields. In that reference
no vector nor axial--vector contributions to the direct vertices in 
$K \rightarrow 2 \pi / 3 \pi$ were found. The scalar and pseudoscalar
resonance contributions are shown in the fifth column of Table 6 where
$k_F$ is the FM factor.
\par
Our non--vanishing $N_i$ combinations contributing to $K \rightarrow 
\pi \pi ( \pi )$ are shown in Table 7. In the just quoted previous works
all the entries in this Table were zero. We get, therefore, 
\underline{new} contributions.

\begin{sidewaystable}
\begin{center}
\begin{tabular}{|c|c|c|c|c|} 
\hline
\multicolumn{1}{|c|}{} & 
\multicolumn{1}{|c|}{} & 
\multicolumn{1}{|c|}{} & 
\multicolumn{2}{|c|}{Phenomenological result} \\  
\multicolumn{1}{|c|}{Counterterm} & 
\multicolumn{1}{|c|}{Vectors} & 
\multicolumn{1}{|c|}{Axial--Vectors} & 
\multicolumn{2}{|c|}{with $\omega_1^R = \sqrt{2} \, \Frac{m_R^2}{F_{\pi}^2}
f_R \, \eta_R$} \\
\cline{4-5}
\multicolumn{1}{|c|}{} & 
\multicolumn{1}{|c|}{} & 
\multicolumn{1}{|c|}{} & 
\multicolumn{1}{|c|}{Vectors} & 
\multicolumn{1}{|c|}{Axial--Vectors} \\ 
\hline
\hline
& & & &  \\
$N_1^r + N_2^r$ & - & $2 \Frac{F_{\pi}^2}{m_A^2} ( \gamma_1 - 2 \gamma_2
- 2 \gamma_4 ) \, \omega_1^A$ & - & $0.007 \eta_A$ \\
& & & &  \\
\hline
& & & & \\
$N_3$ & - & $-2 \Frac{F_{\pi}^2}{m_A^2} (\gamma_1 + \gamma_3) \, 
\omega_1^A$ & - & $-0.001 \eta_A$ \\
& & & & \\
\hline
& & & & \\
$N_9^r$ & $\Frac{F_{\pi}^2}{m_V^2} \, \beta_V \, \omega_1^V$ &  -  & 
$-0.005 \eta_V$ &  -  \\
& & & & \\
\hline
\end{tabular} 
\caption{${\cal O}(p^4)$ counterterms with vector and axial--vector
contributions relevant for $K \rightarrow 2\pi / 3 \pi$. }
\end{center}
\end{sidewaystable}

We find $N_1^V+N_2^V=0$ independent of factorization. This same result
was also found in Refs.~\cite{EK93,IP92} but with  the factorization
hypothesis. Using the phenomenological values for the strong 
coupling of vectors and axials in Tables 1 and 2 
we obtain a determination of the $N_i$  
combinations (see fourth and fifth column in  Table 7) and then to
counterterms contributing to $K\rightarrow \pi\pi (\pi)$~: $K_1, \ K_2$ 
and $K_3$ (fourth column in Table 6).
\par
In our framework we find for the first time a direct (not given by 
$L_i$'s) vector and axial--vector contribution to $K \rightarrow \pi
\pi$ and $K \rightarrow \pi \pi \pi$. In particular 
it is worth noting that $K_1 $ contributes to $A_0$ of  
$K\rightarrow \pi\pi$ (see Eq.~(\ref{eq:k23pi})), which thus gets 
contributions from vectors in this formulation
\footnote{We can split the massive spin--1 field propagator 
${\cal D}_{\mu \nu}(k)$ as \cite{deWit} 
${\cal D}_{\mu \nu} (k) = {\cal D}_{\mu \nu}^s (k) + {\cal D}_{\mu \nu}^t 
(k)$ where `s' stands for space--like and `t' for time--like polarizations.
In this way $k^{\mu} {\cal D}_{\mu \nu}^s (k) = 0$ but  
$k^{\mu} {\cal D}_{\mu \nu}^t (k) = - k_{\nu} / m_V^2$,
with $m_V$ the mass of the field. We interpret the vector contribution
to $K \rightarrow \pi \pi$ as generated by a local term (non--pole)
through the time--like part of the vector propagator. This is analogous to 
the spin--1 $W$-boson contribution to $\pi\rightarrow \mu (e)\nu$.}.
\par
Before we proceed to establish the conclusions we reach from our results,
several cautious remarks concerning the status of the study of the
$K \rightarrow \pi \pi (\pi)$ channels have to be made~: i) the experimental
error in the strong \opc couplings ($L_i$) which also contribute to these
processes has not been taken into account in the analysis of 
Refs.~\cite{EK93,KMW91}; ii) isospin breaking corrections, potentially
relevant, have not been studied yet and iii) experiments with better 
accuracy are needed in order to determine more precisely the \opc weak
counterterm combinations $K_i$ (see a preliminary study in 
Ref.~\cite{serp97}).
\par
Due to these incertitudes one cannot make a conclusive statement or an
accurate quantitative analysis. From our results, however, we think it is
conservative to conclude that~:
\begin{itemize}
\item[a)] Our contributions improve the phenomenological agreement.
If we take a look to our $V+A$ contribution in the fourth column of 
Table 6, add it to the $S+P$ contribution in the fifth column and compare
with the phenomenological values in the last three columns we note that
in the three cases $K_1$, $K_2$ and $K_3$ our contributions aim to 
improve the original result of Ecker et al. In fact a complete 
phenomenological agreement, once all contributions are considered, 
might indicate that $\eta_V > \eta_A$.
\item[b)] As seen in Table 6  there is no splitting between $K_2$ and 
$K_3$ in the FM evaluation of the $S+P$ contributions while the
phenomenological values seem to indicate it. Those contributions coming
from $V+A$ in our framework show a splitting that, however, goes in 
the opposite direction to the phenomenological one. We note that $K_2$
and $K_3$ get also a contribution from $\gamma_3$ and $\gamma_4$ 
in Table 2 that are suppressed by the number of colours expansion and 
unknown (we have used $\gamma_3 = \gamma_4 = 0$ in Tables 6 and 7). It cannot
be excluded that this contribution helps to recover agreement with
the phenomenological value of the splitting. Due to the sensitivity of 
the weak counterterm contributions to $K \rightarrow \pi \pi \pi$ to the
$\gamma_i$ couplings we think that a better control on these couplings
is needed.
\item[c)] From our analysis we see that 
a large $\eta_V$ and $\eta_A$, let us say ${\cal O}$(1) or even
larger, fits better the phenomenology of  $K\rightarrow \pi\pi (\pi )$.
This is not the case, however, for the other decays we have studied,
like for instance  $K^\pm \rightarrow \pi^\pm  \gamma \gamma$
and the anomalous non-leptonic kaon decays \cite{DPN98}, where a value 
close to the Wilson coefficient, i.e. 0.2-0.3, gives better agreement
with the phenomenology. Of course at leading order
the  vector coupling is unique for all decays,
but ${\cal L}_{4}^{|\Delta S|=1} $ is evaluated
with couplings at the scale of the resonance. We may
interpret these $\eta_V \simeq \eta_A \simeq 1$ in $K \rightarrow
\pi \pi (\pi)$, at the phenomenological level,
as generated by a non-perturbative enhancement of the `running' couplings
$\eta_V$ and $\eta_A$ between the scale of the
$\rho$ and the $m_K$, which happens in these particular processes 
($K\rightarrow \pi\pi (\pi )$). The same, we know, is happening for 
$G_8$ (see discussion in Section 2.3).
\end{itemize}

As a conclusion we find that, in contradistinction to what happens in the 
antisymmetric formulation \cite{EK93} or the hidden symmetry \cite{IP92}, 
VMD goes in the direction addressed by the phenomenology if the conventional
vector formulation is employed. To disentangle which one is the right approach
it would be crucial an accurate phenomenological determination of the 
$K_i$ terms and, in particular, of $K_1$.

\subsection{Non-anomalous radiative non-leptonic kaon decays}
\hspace*{0.5cm}
Our formulation gives new insights in several of these processes too.
\vspace*{0.7cm} \\
\underline{${\bf K\rightarrow \pi \gamma^*}$}
\vspace*{0.5cm} \\
$K^\pm\rightarrow \pi^\pm \gamma^*$ and $K_S\rightarrow \pi^0 \gamma^*$
get their leading contribution in $\chi$PT at \opc with loops  
and the following counterterm combinations \cite{EPR87}~:
\begin{eqnarray}
\omega_{+} \, & = & \, \Frac{64 \pi^2}{3} \, \left[ \, N_{14}^r \, - \, 
N_{15}^r \, + \, 3 L_9^r \, \right] \, + \, \Frac{1}{3} \ln \left( 
\Frac{\mu^2}{m_K m_{\pi}} \right)~, \nonumber \\
\omega_S \, & = & \, \Frac{32 \pi^2}{3} \, \left[ \, 2 N_{14}^r \, + 
\, N_{15}^r \, \right] \, + \, \Frac{1}{3} \ln \left( \Frac{\mu^2}{m_K^2} 
\right)~, \label{eq:kpgct}
\end{eqnarray}
where $\mu$ is the renormalization scale and cancels the scale
dependence of the counterterm combinations.
Since we use vector meson saturation $\mu \simeq m_{\rho}$.
\par
No useful experimental information exists on
$K_S\rightarrow \pi^0 \gamma^*$, but DA$\Phi$NE \cite{DE95} will either  
measure the branching ratio or will put an interesting bound,
which it will turn in a non-trivial constrain on $\omega_S$.
$\omega_{+}$ has been measured in the two possible lepton final states:
muon and electron. Here $\omega_{+}$ can be extracted from the rate and
the spectrum. In $K^\pm\rightarrow \pi^\pm e^+ e^-$ the value extracted 
for $\omega_{+}$ from a combined fit of the rate and spectrum is~: 
$0.89^{+0.24}_{-0.14}$ \cite{Alliegro}. If only the rate is considered
the result is $\omega_{+} = 1.20 \pm 0.04$ \cite{Alliegro,BNL78712}.
From the rate of $K^\pm \rightarrow \pi^\pm \mu^+ \mu^-$ the value
$\omega_{+}=1.07\pm0.07$ is obtained \cite{BNL78712}. The measured ratio
$\Gamma(K^\pm\rightarrow \pi^\pm \mu^+ \mu^-)/
\Gamma(K^\pm\rightarrow \pi^\pm e^+ e^-)$ is over 2 $\sigma$'s
away from the \opc $\chi$PT prediction \cite{BNL78712} for a 
wide range of values of $\omega_{+}$, which includes the ones given above.
In our formulation the relations for these counterterms
are independent of factorization (see Eqs.~(\ref{eq:gva},\ref{eq:axials}) 
and Table 4). Also we have a new axial--vector contribution (see Table 4) 
proportional
to $\alpha_A$. With the values in Tables 1 and 2 we obtain
\begin{equation}
\omega_{+} \, = \, 4.4 \, - \, 4.2 \, \eta_V \, + \, 0.8  \, \eta_A
\; \; \; \; \; \; , \; \; \; \; \; \; \;  \; 
\omega_S \, = \, 0.3 \, + \, 8.4 \, \eta_V~.
\label{eq:kpgref}
\end{equation}
Thus we expect some cancellation for $\omega_{+}$ among weak and 
strong vector contributions. A value 
$\eta_V \simeq 1$ could still accommodate the phenomenology
of this channel and the one of $K\rightarrow \pi\pi (\pi)$, 
as commented in the previous Subsection.
However such solution might be  in contradiction
with the value of $\eta_V $, which can be deduced, as we shall see, from
$K^\pm \rightarrow \pi^\pm  \gamma \gamma$ and from the anomalous 
non-leptonic kaon decays (see also Ref.~\cite{DPN98}).
\par
We think it is premature a final conclusion on the value of 
$\omega_{+}$ due to the possible disagreement among the
\opc $\chi$PT spectrum
and experiments. As it happens in $K_L\rightarrow \pi^+\pi^- \gamma$ 
\ops contributions could be important \cite{DPN98} (see the
discussion on \ops corrections to $G_8$ in Ref.~\cite{DG95}).
\par
From the expression of $\omega_S$ in Eq.~(\ref{eq:kpgref}) we notice
that $K_S\rightarrow \pi^0 \ell^+ \ell^-$ is very sensitive to 
the value of $\eta_V$. In Table 8 we show the predictions for the 
branching ratio for three representative values of $\eta_V$. DA$\Phi$NE 
should be crucial for a better
phenomenological understanding of this process since it should be able
to measure $Br(K_S \rightarrow \pi^0 e^+ e^-) \, > \, 5 \times  10^{-10}$
and $Br(K_S \rightarrow \pi^0 \mu^+ \mu^-) \, > \, 10^{-10}$ \cite{DE95}.

\begin{table}
\begin{center}
\begin{tabular}{|c|c|c|}
\hline
$\eta_V$ & $Br(K_S \rightarrow \pi^0 e^+ e^-)$ &
$Br(K_S \rightarrow \pi^0 \mu^+ \mu^-)$ \\
\hline
\hline
0.3 & 1.8 $\times \, 10^{-8}$ & 3.9 $\times \, 10^{-9}$  \\
\hline
0.5 & 5.1 $\times \, 10^{-8}$ & 1.1 $\times \, 10^{-8}$  \\
\hline
1.0 & 2.1 $\times \, 10^{-7}$ & 4.4 $\times \, 10^{-8}$  \\
\hline
\end{tabular}
\caption{Branching ratios of $K_S \rightarrow \pi^0 \ell^+ \ell^-$ 
($\ell = e, \mu$) for different values of $\eta_V$ using $\omega_S$ 
in Eq.~(\ref{eq:kpgref}).}
\end{center}
\end{table}

In Ref.~\cite{EK93} a different approach has been proposed.
Using procedure \pra of Section 2.3
for the counterterm combinations in Eq.~(\ref{eq:kpgref}) they obtain
\footnote{We note that there is a numerical difference  
between the model independent contribution to $\omega_{+}$ in 
Eqs.~(\ref{eq:kpgref},\ref{eq:kpgrefEK93}) and also
the factor multiplying the parentheses in Eq.~(\ref{eq:omegasplus}) (below)
and the analogous one in Eq.~(\ref{eq:kpgrefEK93}). This is due to the fact
that we use different values for the $f_V$ coupling.} \cite{DE95}~:
\begin{eqnarray}
\omega_{+} & = & 5.3 \, - \, 6.9 k_F \, - \, \Frac{(16\pi )^2}{3} k_F 
H_1^r (m_\rho)~, \nonumber \\
\omega_S & = & 0.3 \, + \, 2.3 k_F - \, \Frac{(16\pi )^2}{3} k_F 
H_1^r (m_\rho)~, \nonumber \\
\omega_{S} & = & \omega_{+} \, + \, 4.6 \, ( \, 2 k_F \, - \, 1 \, ) \, - \, 
0.43~. 
\label{eq:kpgrefEK93}
\end{eqnarray}
If we  use VMD for $H_1$
the same result than ours in Eq.~(\ref{eq:kpgref}) would be obtained 
(with $\eta_V = \eta_A = k_F$ and neglecting the new axial contribution 
proportional
to $\alpha_A$) \footnote{VMD gives $H_1^r (m_\rho) = -(f_V^2 + f_A^2)/8$.
This result is consistent with the evaluation of $H_1$ in the ENJL model 
\cite{BBR93}.}. However in Ref.~\cite{EK93} the choice 
$H_1^r(m_{\rho})=0$ has been advocated. 
This is a clear departure from VMD and must be considered a local contribution.
Of course, if the suggestion $H_1=0$ is supported by the
phenomenology, we can implement it 
in our scheme by adding a suitable local contribution.
Notice that the last relation in Eq.~(\ref{eq:kpgrefEK93}) is independent
of the $H_1$ coupling and arises in procedure ${\cal A}$ as a consequence
of factorization. In our formulation this relation translates into
\begin{equation}
\omega_S \, = \, \omega_{+} \, + \, 6.3 \, ( \, 2 \, \eta_V \, - \, 1 \, 
- \, 0.12 \, \eta_A \, ) \, - \, 0.43~,
\label{eq:omegasplus}
\end{equation}
but now it is independent of factorization, hence we conclude that
theirs too. Also chiral symmetry
tells us that a possible additional local term, if any, generated by
$H_1$ (to be included into $\omega_{+}$ and $\omega_S$) cancels in this
relation.
\par
Our insight in these channels is to have clarified that the previous 
results (that coincide with ours if we assume VMD and neglect axial--vector
contributions) are model--independent and do not rely on factorization
(see for instance Eq.~(\ref{eq:omegasplus})).
\vspace*{0.7cm} \\
\underline{${\bf K^\pm \rightarrow \pi^\pm  \gamma \gamma}$}
\vspace*{0.5cm} \\
Due to the recent BNL measurement \cite{TAK96}  
this channel is now particularly interesting.
It starts at \opc in $\chi$PT with a finite loop contribution and the scale 
independent counterterm combination \cite{EPR88}
\begin{equation}
\hat{c} \; = \; \Frac{128 \, \pi^2}{3} \, \left[ \, 3 ( L_9 + L_{10}) \, 
+ \, N_{14} \, - \, N_{15} \, - \, 2 \, N_{18} \, \right] \; \; \; ,
\label{eq:chat}
\end{equation}
where $L_9$ and $L_{10}$ are couplings in the \opc $\chi$PT 
strong lagrangian ${\cal L}_4$ \cite{GL85}. There is no complete study at 
\ops but the
unitarity corrections from $K \rightarrow \pi \pi \pi$ have been computed
\cite{DP96} and enhance the \opc rate by $30-40\%$. Vector exchange at 
this order has been proven to be negligible \cite{DPN97,DP96}. 
The BNL measurement in fact has a better fit with these \ops
contributions \cite{TAK96}. At \opc the 
vector contribution cancels in the strong ($L_i$) and in the weak sector 
($N_i$) (see Table 4). In the formulation used in this paper this result
is independent of factorization. Using 
the values in Tables 1 and 2 and the expressions in Table 4
we get 
\begin{equation}
\hat{c} \, = \, 2.15 \, - \, 4.2 \, \eta_A~.
\label{eq:chatn}
\end{equation}
Thus $\hat{c}$ is sensitive to the weak coupling of axials
and hence $\alpha_A$ (see Table 4) might correct the result in 
Ref.~\cite{EK93}.
We observe that the recent experimental figure $\hat{c}=1.8\pm 0.6 $ 
\cite{TAK96} (this is the value obtained when unitarity corrections
have been included, as supported by a better $\chi^2$) prefers a small but
non--vanishing value $\eta_A \simeq 0.2-0.3$. Thus the perturbative evaluation
seems consistent with data. Even a slight improvement in the actual
experimental situation would fix unambiguously $\eta_A$.
\vspace*{0.7cm} \\
\underline{${\bf K \rightarrow \pi \pi (\pi)  \gamma }$}
\vspace*{0.5cm} \\
In the amplitudes of these processes one generally 
separates the  bremsstrahlung contribution,
which can be related to the non-radiative amplitude
by the Low Theorem \cite{DE97,Low}, and
the direct emission (structure dependent) component to 
$K \rightarrow \pi \pi (\pi)  \gamma $
\cite{DE95,TONI95,DAGI96}. From the spectrum in the photon energy the 
two contributions can be distinguished. 
\par
The direct emission amplitudes, according to their transformation under
parity, are divided in electric and magnetic contributions \cite{DMS92}
and get their first non--vanishing contributions at \opc in $\chi$PT.
The magnetic terms come from odd--intrinsic parity terms and will be 
commented in the next Subsection. 
\par
Only one \opc weak counterterm  combination
appears for $K \rightarrow \pi \pi   \gamma $ decays:
 $N_{14} \,-\, N_{15} \,-\,
N_{16} \,-\,  N_{17}$ (see Table 9) and it is 
scale independent \cite{Esposito90,DI95,DMS93,EN92}.
Thus the chiral loop contribution is finite.
As we can see from Tables 4 and 9, we have an almost complete cancellation
of vector contributions, independently of factorization.
Our axial--vector contributions (and consequently the bulk of the total
contribution) are about $50 \%$ bigger than the result of
Ref.~\cite{EK93}, due to our different results  for
$N_{16}^A $ and $ N_{17}^A$.
\par
The electric amplitudes for $K_L(K^\pm) \rightarrow \pi \pi \pi   \gamma $ 
also get the previous counterterm combination \cite{DE97}. Meanwhile
for $K_S \rightarrow \pi^+ \pi^-\pi^0 \gamma $ the relevant \opc scale 
dependent counterterm combination is
$
7(N_{14}^r \,-\, N_{16}^r )\,+\,5( N_{15}^r \,+\,  N_{17})
\, \simeq \,  0.39 \, \eta_V \, + \, 0.01 \, \eta_A
$
where we have quoted our spin--1 exchange contributions.
Our value is larger ($30\%$) than the one obtained with
factorization in the antisymmetric formulation \cite{EK93,DE97} due to 
our different result for $N_{16}^A$ and $N_{17}^A$.
\par
In Table 9, the full list of \opc VMD  contributions is reported.
Our results will differ particularly from
Ref.~\cite{EK93}, when axials are important and $N_{16}^A$
and $N_{17}^A$ are involved. This happens, for instance, in 
the channel $K_L\rightarrow \pi^+ \pi^- \gamma^*$, which is relevant now due
the  preliminary data from Fermilab \cite{Win97}, which for the
first time measures this decay.
This \opc electric contribution is in competition with 
the bremsstrahlung CP violating and the magnetic amplitudes.
\par
For a thorough review on these processes see the 
Refs.~\cite{DE95,TONI95,DAGI96,DE97,DI95}.

\begin{table}
\begin{center}
\begin{tabular}{|c|c|c|} 
\hline
\multicolumn{1}{|c|}{} & 
\multicolumn{1}{|c|}{} & 
\multicolumn{1}{|c|}{Phenomenological result} \\  
\multicolumn{1}{|c|}{Counterterm combination} &
\multicolumn{1}{|c|}{Processes} &
\multicolumn{1}{|c|}{with $\omega_1^R = \sqrt{2} \, \Frac{m_R^2}{F_{\pi}^2}
f_R \, \eta_R$, $\;$  $\omega_2^R = - \omega_1^R$} \\
\hline
\hline
$N_{14}^r - N_{15}^r$ & $K^+ \rightarrow \pi^+ \gamma^*$ & \\
& $K^+ \rightarrow \pi^+ \pi^0 \gamma^*$ & $-0.020 \, \eta_V \, + \, 
0.004 \, \eta_A$ \\ 
\hline
$2 N_{14}^r + N_{15}^r$ & $K_S \rightarrow \pi^0 \gamma^*$ & \\
& $K_L \rightarrow \pi^0 \pi^0 \gamma^*$ & $0.08 \, \eta_V$ \\ 
\hline
$N_{14} - N_{15} - 2 N_{18}$ & $K^+ \rightarrow \pi^+ \gamma \gamma$ & \\
& $K^+ \rightarrow \pi^+ \pi^0 \gamma \gamma$ &  $-0.01 \, \eta_A$ \\
& $K_S \rightarrow \pi^+ \pi^- \gamma \gamma$ & \\
\hline
$N_{14} - N_{15} - N_{16} - N_{17}$ & $K^+ \rightarrow \pi^+ \pi^0 \gamma$ 
& \\
& $K_S \rightarrow \pi^+ \pi^- \gamma$ & \\
& $K^+ \rightarrow \pi^+ \pi^+ \pi^- \gamma$ & $0.002 \, \eta_V \, - \, 
0.010 \, \eta_A$ \\
& $K^+ \rightarrow \pi^+ \pi^0 \pi^0 \gamma$ & \\
& $K_L \rightarrow \pi^+ \pi^- \pi^0 \gamma$ & \\
\hline 
$7(N_{14}^r-N_{16}^r) + 5 (N_{15}^r+N_{17})$ & $ K_S \rightarrow \pi^+
\pi^- \pi^0 \gamma$ & $ 0.39 \, \eta_V \, + \, 0.01 \, \eta_A$ \\
\hline
$N_{14}^r - N_{15}^r - 3( N_{16}^r - N_{17}) $ & $ K_L \rightarrow \pi^+ \pi^- 
\gamma^*$ & $ -0.004 \, \eta_V \, + \, 0.018 \, \eta_A$ \\
\hline 
$N_{14}^r - N_{15}^r - 3( N_{16}^r + N_{17}) $ & $ K_S \rightarrow \pi^+ \pi^- 
\gamma^*$ & $0.05 \, \eta_V \, - \, 0.04 \, \eta_A$ \\
\hline
$N_{14}^r + 2 N_{15}^r -3(N_{16}^r-N_{17})$ & $ K^+ \rightarrow \pi^+ \pi^0 
\gamma^*$ & $ 0.12 \, \eta_V \, + \, 0.01 \, \eta_A$ \\
\hline
$N_{29} + N_{31}$ & $ K_L \rightarrow \pi^+ \pi^- \gamma$ & \\
& $K^+ \rightarrow \pi^+ \pi^+ \pi^- \gamma$ & $0.005 \, \eta_V \, + \, 
0.003 \, \eta_A$ \\
& $ K_S \rightarrow \pi^+ \pi^- \pi^0 \gamma$ & \\
\hline 
$ 3 N_{29} - N_{30}$ & $ K^+ \rightarrow \pi^+ \pi^0 \gamma$ & \\
& $K^+ \rightarrow \pi^+ \pi^0 \pi^0 \gamma$ & $ -0.005 \, \eta_V \, - \, 
0.003 \, \eta_A$ \\
\hline
$5 N_{29} - N_{30} + 2 N_{31}$ & $ K_S \rightarrow \pi^+ \pi^- \pi^0 
\gamma$ & $ 0.005 \, \eta_V \, + \, 0.003 \, \eta_A$ \\
\hline
$6 N_{28} + 3 N_{29} - 5 N_{30}$ & $ K_L \rightarrow \pi^+ \pi^- \pi^0 
\gamma$ & $ - 0.004 \, \eta_V \, - \, 0.003 \, \eta_A$ \\
\hline
\end{tabular} 
\caption{Counterterm combinations appearing in radiative non--leptonic
kaon decays and the vector and axial--vector contribution to them. 
The assumption of factorization ($\omega_2^R = - \omega_1^R$) has been 
used only in the terms involving $N_{28},$ ...$N_{31}$. For the numerical
results we use the experimental value of the couplings, when available,
and the ENJL predictions in Tables 1 and 2. However in the last 
counterterm combination we notice that the coefficient of the vector
contribution is very sensitive to the value of the $\theta_V$ coupling
chosen, i.e. if we had chosen $\theta_V = 2 h_V$ (HS prediction)
the coefficient would be $+ 0.037$ instead of $-0.004$. This is the only 
entry where the two model predictions for the strong 
couplings in Tables 1 and 2 give a substantially different result.}
\end{center}
\end{table}

\subsection{Anomalous radiative non-leptonic kaon decays}
\hspace*{0.5cm} 
In our spin-1 formulation the weak couplings associated to anomalous
transitions ($N_{28}, ...N_{31}$) are particularly interesting. 
Contrarily to the antisymmetric formulation \cite{EK93}, we do get 
vector and axial--vector contributions to these counterterms (see Table 5).
This should be interpreted in the same terms as the failure of
the antisymmetric formulation of the vectors to describe the 
strong lagrangian $V \rightarrow P \gamma$ (see discussion in  section 2.1)
at the proper chiral order. 
\par
Due to previous studies in factorization \cite{CH90} it has become 
customary to rewrite these couplings as 
\begin{eqnarray}
a_1 \, = \, 8 \, \pi^2 \, N_{28} \; \; \; \; \;  & \; , \; & \; \; \; \; \; \; 
a_2 \, = \, 32 \, \pi^2 \, N_{29} \; , \nonumber \\
a_3 \, = \, \Frac{16}{3} \, \pi^2 \, N_{30} \; \; \; \; \; & \; , \; & 
\; \; \; \; \; \; 
a_4 \, = \, 16 \, \pi^2 \, N_{31} \; . 
\label{eq:aidef}
\end{eqnarray} 
The $a_i$ are positive parameters of ${\cal O}$(1). 
As pointed out in relation with the Factorization Model in Subsection 2.3
these couplings get a factorizable contribution due the Wess--Zumino--Witten
anomaly action \cite{EN92}. If factorization works this contribution would
be the same for all $a_i$'s in Eq.~(\ref{eq:aidef}) that we call
$a_i = \eta_{FM}$, $i=1,2,3,4$. To these we should add the vector and 
axial--vector contributions that we have found in our work.
\par
The octet operators $W_{28},..., W_{31}$ are relevant for several processes:
$K\rightarrow \pi \pi \gamma$ \cite{ENP94,DPN98},
$K\rightarrow \pi \pi \pi \gamma$ \cite{DE97,ENP94}, etc. 
For  the complete analysis of 
$K_L\rightarrow \pi^+ \pi^- \gamma$ at \ops we refer to our work 
\cite{DPN98}.
Regarding the $K^\pm \rightarrow \pi^\pm \pi^0 \gamma$ process,
if one neglects higher order contributions, one can write the \opc amplitude
as \cite{CH90}
\begin{equation}
A_4 \; \equiv \;  - \, ( \, 2 \, - \, 3 \, a_2 \, + \, 6 \, a_3)  \; 
\simeq \; 
- \, ( \, 2 \, + \, 3 \, \eta_{FM} \, + \, 1.7 \, \eta_V \, + \, 0.9 \, 
\eta_A)~.
\label{eq:k2pipg}
\end{equation}
The measured branching ratio \cite{PDG} implies (neglecting
\ops corrections) $A_4=-4.5\pm 0.5$
\cite{ENP94}. Thus our new contributions go again in the right direction.
\par
We could conclude that the phenomenology of the aforementioned decays
prefers a small value of the factorization factors ($\eta_{FM}$, 
$\eta_V$, $\eta_A$), i.e., inside the errors, they can be identified
with the Wilson coefficient in Eq.~(\ref{eq:wilson}).
\par
We refer to Table 9 for a complementary list of the counterterm 
contributions relevant for other radiative non--leptonic kaon decays and
our predictions for them.

\section{Conclusions}
\hspace*{0.5cm}
We have given a complete description of the spin-1 meson exchange 
contributions to the \opc weak chiral lagrangian in the conventional
vector formulation. A detailed comparison with the results obtained in the  
antisymmetric formulation \cite{EK93} has also been developed.
\par
In the comparison we find the following observations and new results~:
\begin{itemize}
\item[i)] In our formulation the leading
weak lagrangian for the vectors (see Eq.~(\ref{eq:vpp1})) is rather 
constrained, while the  strong sector 
(see Eqs.~(\ref{eq:evf},\ref{eq:eaf})) is very rich. This is at odds
with the antisymmetric formulation, where the opposite is true:
constrained strong lagrangian and rich weak lagrangian.
This is due to the relatively different chiral power at which both 
sectors (strong and weak) get their leading contribution. Since we know 
better (from a phenomenological point of view) the strong and electromagnetic
decays of vector mesons, we think that our
procedure is in better shape to give more information with less
assumptions. In this work and when necessary we have relied on
models (typically ENJL \cite{PradesZ}) to get the strong couplings, but 
these are more under control than the weak couplings of vectors  
and in any case experiments
should be able to determine the full strong  sector in the near future.
Due to our very restrictive weak lagrangian for vectors and axial--vectors, 
all the results obtained in factorization in the antisymmetric
formulation are obtained by us but without the use of
factorization (for $N_{16}^A$ and $N_{17}^A$ see Section 3).
We think that this is due to the fact that chiral symmetry and Lorentz
invariance imposes to the ${\cal O}(p)$ weak lagrangian linear in the
resonance fields ${\cal L}_R^{{\cal O}(p)}$ in Eq.~(\ref{eq:vpp1})
the factorizable structure not as a hypothesis of work but as a 
consequence of the symmetry and therefore model--independent.
We have got new contributions and shown their phenomenological relevance.
This gives  a more solid ground to these results
and complements the analysis of Ref.~\cite{EK93}.
\item[ii)] We emphasize that our results for $K \rightarrow \pi \pi 
(\pi)$ and non--anomalous radiative decays are completely model--independent.
The hypothesis of factorization only enters (and with a very minor r\^ole)
in the processes involving $N_{28},...,N_{31}$, and therefore anomalous
radiative kaon decays. Due to our very simple, constrained and also general 
procedure (factorization is used just to relate $N_{30}^V$ to the
other $N_i^V$ and $N_{31}^A$ to the other $N_i^A$), we parameterize
all the spin-1 resonance contributions to \opc weak 
lagrangian in terms of only two physical weak couplings of vectors and
axials:  $\eta_V$ and $\eta_A$ (see Table 9).
\item[iii)] We obtain non--vanishing vector and axial--vector exchange
contributions in odd-intrinsic parity terms. These appear only at
higher orders in the antisymmetric formulation.
This problem was already realized in the strong sector for the 
$VP\gamma$ vertex, contributing at \ops to $\gamma\gamma\rightarrow
\pi^0\pi^0$ in the vector formulation
and ${\cal O}(p^8)$ in the antisymmetric
formulation, in contradiction with QCD properties \cite{EP90}. 
\item[iv)] Regarding the \opc axial--vector contributions to
$N_{16}$ and $N_{17}$ we differ from the ones in Ref.~\cite{EK93}
and this is phenomenologically relevant.
\item[v)] We get new contributions to the  $N_{i}$, due to our richer strong
lagrangian: all the terms  proportional to $\alpha_{V,A}\ ,\
\beta_V, \ $ and $\gamma_i$ are new. The ones to $K\rightarrow
\pi\pi (\pi)$ seem to us of immediate phenomenological impact
and thus here we have made a more detailed study of the phenomenological
implications. 
\item[vi)] We have also commented that in an explicit analysis
of $K_L\rightarrow \pi^+\pi^- \gamma$, where
\opc and \ops  VMD contributions were simultaneously present,
our approach and the vector realization  give a
theoretical consistent and also phenomenologically successful
picture. Actually it is shown that our treatment of weak \opc VMD is 
compulsory in order to have the correct description of  weak \ops VMD
\cite{DPN98}.
\end{itemize}

Experiments should decide which of the formulations is better. Wherever 
this is not possible we suggest to choose the one which is able to 
describe more physics at the lowest order such to have a better convergence. 
\vspace*{1cm} \\
{\bf Acknowledgements}
\vspace*{0.5cm} \\
The authors wish to thank F.J. Botella, F. Cornet and G. Isidori
for interesting and fruitful discussions on the
topics of this paper. J.P. is partially supported by Grant PB94--0080 of
DGICYT (Spain) and Grant AEN--96/1718 of CICYT (Spain).

\newpage

\appendix
\newcounter{homero}
\renewcommand{\thesection}{\Alph{homero}}
\renewcommand{\theequation}{\Alph{homero}.\arabic{equation}}
\setcounter{homero}{1}
\setcounter{equation}{0}

\section*{Appendix A~: The weak $\omega_i^R$ couplings in the Factorization 
Model in the Vector couplings}
\hspace*{0.5cm}
The $\omega_i^R$ weak couplings defined in Eq.~(\ref{eq:vpp1}) for 
$R= V,A$ can be evaluated in the FMV we have proposed in Ref.~\cite{DPN97}.
The bosonization of the $Q_{-}$ operator in Eq.~(\ref{eq:qplusm})
can be carried out in the FMV from the strong action $S$ of a chiral 
gauge theory. If we split the strong action and the left--handed 
current into two pieces~: $S = S_1 + S_2$ and ${\cal J}_{\mu} = 
{\cal J}_{\mu}^1 + {\cal J}_{\mu}^2$, respectively, the $Q_{-}$ operator
is represented, in the factorization approach, by
\begin{equation}
Q_{-} \, \leftrightarrow \, 
4 \, \left[ \, \langle \, \lambda \, \{ {\cal J}_{\mu}^1 \, , \, 
{\cal J}^{\mu}_2 \, \} \, \rangle \, - \, \langle \, \lambda
\, {\cal J}_{\mu}^1 \, \rangle \, \langle \, {\cal J}^{\mu}_2 \, 
\rangle \, - \, \langle \, \lambda \, {\cal J}_{\mu}^2 \, \rangle \, 
\langle \, {\cal J}^{\mu}_1 \, \rangle \, \right]~,
\label{eq:a1}
\end{equation}
with $\lambda \equiv ( \lambda_6 - i \lambda_7) /2$ and, for generality,
the currents have been supposed to have non--zero trace.
\par
In order to apply this procedure to construct the factorizable contribution
to the weak ${\cal O}(p)$ lagrangian in Eq.~(\ref{eq:vpp1}) we have to 
identify in the full strong action the pieces that can contribute at this
chiral order. We define, correspondingly,
\begin{equation}
S \,  =  \, S_{R \gamma} \, + \, S_2^{\chi}~, 
\label{eq:a2}
\end{equation}
where the actions correspond to the lagrangian densities proportional to 
$f_V$ in ${\cal L}_V$ (Eq.~(\ref{eq:evf})) or to $f_A$ in ${\cal L}_A$
(Eq.~(\ref{eq:eaf})) ($S_{R \gamma}$) and ${\cal L}_2$ in Eq.~(\ref{eq:str2})
($S_2^{\chi}$).
\par
Evaluating the left--handed currents and keeping only the terms of 
interest we get
\begin{eqnarray}
\Frac{\delta \, S_{R \gamma}}{\delta \, \ell^{\mu}} \, & = & \, 
- \, \Frac{f_R}{\sqrt{2}} \, m_R^2 \, u^{\dagger} \, R_{\mu} \, u~,
\nonumber \\
\Frac{\delta \, S_2^{\chi}}{\delta \, \ell^{\mu}} \, & = & \, 
- \, \Frac{F_{\pi}^2}{2} \, u^{\dagger} \, u_{\mu} \, u~.
\label{eq:a3}
\end{eqnarray}
Then the effective action in the factorization approach is
\begin{eqnarray}
{\cal L}_R^{fact} \, & = & \, 4 \, G_8 \, \eta_R \, 
\left[ \, \langle \, \lambda \, \left\{ 
\Frac{\delta S_{R\gamma}}{\delta \ell^{\mu}} \,
, \, \Frac{\delta S_{2}^{\chi}}{\delta \ell_{\mu}} \, \right\} \rangle
\, - \, \langle \lambda \Frac{\delta S_{R\gamma}}{\delta \ell^{\mu}} \rangle
\langle \Frac{\delta S_{2}^{\chi}}{\delta \ell_{\mu}} \rangle \right.
\nonumber \\
& & \; \; \; \; \; \; \; \; \; \; \; \; \; \; \; \left.
\, - \, \langle \lambda \Frac{\delta S_{2}^{\chi}}{\delta \ell^{\mu}} 
\rangle \langle \Frac{\delta S_{R\gamma}}{\delta \ell_{\mu}} \rangle \, 
\right] 
\; + \, h.c. \; ,
\label{eq:a4} 
\end{eqnarray}
and identifying with ${\cal L}_R^{{\cal O}(p)}$ in Eq.~(\ref{eq:vpp1})
we read
\begin{equation}
\omega_1^R \, = \, \sqrt{2} \,  \Frac{m_R^2}{F_{\pi}^2} \, 
f_R \, \eta_R \; , \; \; \; \; \; \; \; \; \; \; \; \; \; \; \; 
\omega_2^R \, = \, -  \, \omega_1^R \; , 
\label{eq:a5}
\end{equation}
for $R= V$ and $A$.

\end{document}